\documentclass[]{aa}

\usepackage{graphicx}
\usepackage{longtable}
\usepackage{booktabs}
\usepackage{subcaption}
\usepackage[dvipsnames]{xcolor}
\usepackage{placeins}
\usepackage{epstopdf}
\usepackage{tikz}
\usetikzlibrary{shapes.geometric, arrows}
\usepackage[hidelinks,colorlinks=true,linkcolor=blue,urlcolor=blue,citecolor=blue]{hyperref}
\usepackage{xcolor}
\usepackage[normalem]{ulem}

\newcommand{\addedAPA}[1]{\textcolor{blue}{#1}}

\tikzstyle{startstop} = [rectangle, rounded corners, minimum width=3cm, minimum height=1cm,text centered, draw=black, fill=red!30]
\tikzstyle{process} = [rectangle, minimum width=3cm, minimum height=1cm, text centered, draw=black, fill=orange!30]
\tikzstyle{arrow} = [thick,->,>=stealth]

\usepackage{txfonts}

\begin{document}

   \title{A gradient boosting and broadband approach to finding Lyman-$\alpha$ emitting galaxies beyond narrowband surveys}

\titlerunning{A gradient boosting and broadband approach to finding LAE galaxies beyond NB surveys}
   \author{A. Vale
          \inst{1,2}
          \and
          A. Paulino-Afonso\inst{1}
          \and
          A. Humphrey\inst{1,3}
          \and
          P.A.C. Cunha\inst{1,2}
          \and
          B. Ribeiro\inst{4}
          \and
          B. Cerqueira\inst{1,2}
          \and
          R. Carvajal\inst{5,6}
          \and
          J. Fonseca\inst{1,2}
          }
    \authorrunning{Vale et al.}

\institute{Instituto de Astrofísica e Ciências do Espaço, Universidade do Porto, CAUP, Rua das Estrelas, PT4150-762 Porto, Portugal
\and
Departamento de Física e Astronomia, Faculdade de Ciências, Universidade do Porto, Rua do Campo Alegre 687, PT4169-007 Porto, Portugal
\and
DTx–Digital Transformation CoLab, Building 1, Azurém Campus, University of Minho, PT4800-058 Guimarães, Portugal
\and
Celfocus, Avenida Dom João II, 34, Parque das Nações, 1998-031, Lisbon, Portugal
\and
Instituto de Astrofísica e Ciências do Espaço, Universidade de Lisboa, OAL, Tapada da Ajuda, 1349-018 Lisbon, Portugal
\and
Departamento de Física, Faculdade de Ciências, Universidade de Lisboa, Edifício C8, Campo Grande, 1749-016 Lisbon, Portugal}

  \abstract
   {The identification of Lyman-$\alpha$ emitting galaxies (LAEs) has traditionally relied on dedicated surveys using custom narrowband filters, which constrain observations to specific narrow redshift intervals, or on blind spectroscopy, which although unbiased, typically requires extensive telescope time. This makes it challenging to assemble large statistically robust galaxy samples. With the advent of wide-area astronomical surveys producing datasets that are significantly larger than traditional surveys, the need for new techniques arises.}
   {We test whether gradient-boosting algorithms, trained on broadband photometric data from traditional LAE surveys, can efficiently and accurately identify LAE candidates from typical star-forming galaxies at similar redshifts and brightness levels.}
   {Using galaxy samples at $z \in [2,6]$ derived from the COSMOS2020 and SC4K catalogs, we trained gradient-boosting machine-learning algorithms (LGBM, XGBoost, and CatBoost) using optical and near-infrared broadband photometry. To ensure balanced performance, the models were trained on carefully selected datasets with similar redshift and i-band magnitude distributions. Additionally, the models were tested for robustness by perturbing the photometric data using the associated observational uncertainties.}
   {Our classification models achieved F1-scores of $\sim 87\%$ and successfully identified about $7,000$ objects with an unanimous agreement across all models. This more than doubles the number of LAEs identified in the COSMOS field compared with the SC4K dataset. We managed to spectroscopically confirm 60 of these LAE candidates using the publicly available catalogs in the COSMOS field.}
   {These results highlight the potential of machine learning in efficiently identifying LAEs candidates. This lays the foundations for applications to larger photometric surveys, such as Euclid and LSST. By complementing traditional approaches and providing robust preselection capabilities, our models facilitate the analysis of these objects. This is crucial to increase our knowledge of the overall LAE population.}

   \keywords{galaxies: high redshift --
                galaxies: photometry -- surveys -- methods: data analysis -- methods: statistical
               }

   \maketitle
%

\section{Introduction}

\hspace{0.3cm} The Lyman-$\alpha$ (hereafter, Ly$\alpha$) emission line is one of the most prominent spectral features emitted by star-forming galaxies \citep[e.g.][]{PartridgeandPeables,ouchi_2003,dijkstra_2014}. With its short wavelength (1216$\AA$) and intrinsic brightness, the Ly$\alpha$ emission proves highly effective for determining faint high-redshift objects through optical and near-infrared (NIR) observations {\citep[e.g.][]{ouchi_2003,sobral_2018,yuichi_2018}. Ly$\alpha$ emitters (hereafter referred to as LAEs) are galaxies with a detected Ly$\alpha$ emission line with a rest-frame equivalent width ($EW_0$) that exceeds approximately $20$\AA\ {\citep[e.g.][]{shibuya_2019,ReviewLya}}. LAEs are compact galaxies \citep[$r_e \sim 1$ kpc, e.g.][]{ana_afonso_2018}, generally with a low stellar mass \citep[$M_* \sim 10^{8-9} M_\odot$, e.g.][]{pentericci_2007}  and low metallicities \citep[$Z < 2 \times 10^{-2} Z_\odot$, e.g.][]{ReviewLya}. They exhibit an inverse correlation between galaxy size and Ly$\alpha$ emission, where smaller galaxies tend to have larger escape fractions \citep[][]{law_2012,ana_afonso_2018}. Additionally, they present star formation rates of $\sim 1-10$ M$_\odot$ yr$^{-1}$ \citep[e.g.][]{gawiser_2007} and young stellar ages \citep[$\sim 10$ Myr, e.g.][]{hagen_2014, nakajima_2012}. These properties suggest that LAEs trace the early galaxy evolution \citep{taniguchi_2003}.

The most common method for detecting LAEs is narrowband (NB) imaging {\citep[e.g.][]{nb_example_2,nb_example_1}}, which involves comparing images taken using a NB filter (typically 100–200\AA\ wide) {\citep[e.g.][]{NB_technique_1, NB_technique_2}} with a broadband image to identify a flux excess in the NB image. This technique is limited, however, because it only probes specific narrow redshift ranges and is biased toward higher equivalent widths \citep[e.g.][]{matthee_2015}. Furthermore, there is a non-negligible possibility that these {NB-selected samples may include galaxies at lower redshifts with additional emission lines that may be confused with Ly$\alpha$}, such as CIV emission at 1549Å {\citep[e.g.][]{Contam_1}}, MgII at 2798Å {\citep[e.g.][]{ReviewDunlop}}, [OII] at 3727Å {\citep[e.g.][]{Contam_2}},  or [OIII] at 5007Å {\citep[e.g.][]{Contam_3}}. Blind spectroscopy \citep[e.g.][]{slitless} is another technique that is used to search for LAEs. Candidates are identified based on their emission lines without prior photometric selection. While it allows for the detection of LAEs without selection biases, its efficiency is limited by the fact that it often requires extensive telescope time. 
Integral field spectrographs (IFS) can also be used \citep[e.g.][]{ifs_example}. They offer the advantage of capturing the spectra of unbiased galaxy samples. Their effectiveness as wide-area survey instruments is limited by their relatively small field of view, however.\par

{Despite the success of these techniques, they each have different limitations, and as LAE searches expand with upcoming surveys, such as the Large Synoptic Survey Telescope \citep[LSST;][]{lsst} and Euclid \citep{Euclid}, machine-learning methods will be essential to optimize the sample selection and mitigate biases {\citep[e.g.][]{future_surveys}}}. Although the application of machine learning to LAE studies remains limited, some key works have been conducted. \citet{Runnholm_2020} used a linear regression to predict the Ly$\alpha$ luminosity based upon observed properties such as broadband luminosities or other nebular lines. They also used derived physical properties, such as the stellar mass or star formation rate (SFR). \citet{ono_2021} trained convolutional neural network models using simulated images of LAEs to remove contaminants from photometrically selected LAE candidates by performing an image classification based on NB and broadband filter data. Finally, \citet{lorenzo_classifier} developed a machine-learning model based on random forest to select LAEs. They achieved an accuracy of (80 ± 2)\% and a precision of (73 ± 4)\% in the redshift range $z \in$  $[2.5, 4.5]$. At higher redshifts ($z \in$  $[4.5, 6]$), they obtained an accuracy of 73\%  and a precision of 80\%. This model was built on the basis of the physical (stellar mass, SFR, reddening, metallicity, and age) and morphological properties (Sérsic index, half-light radius, and projected semimajor axis), and the training sample was based on several samples of spectroscopically confirmed LAEs from the Great Observatories Origins Deep Survey South field \citep[GOODS-S;][]{goods}, Ultra Deep Survey \citep[UDS;][]{uds}, and Cosmic Evolution Survey Deep \citep[COSMOS;][]{cosmos_2007}.\par

Instead of using derived physical properties, we aim to identify LAEs in a more efficient and cost-effective manner by using only broadband photometric data. We trained supervised machine-learning gradient-boosting algorithms on about 4,000 LAEs from the SC4K survey \citep{sc4k} and non-LAEs (nLAEs) from COSMOS2020 \citep{Cosmos2020} {at $z \in [2,6]$}. We also built on correlations between LAEs and their fluxes, magnitudes, and colors in the optical and NIR broadband information to pave the way for a future application to larger surveys. This is valuable for surveys that allow blind spectroscopy, such as Euclid \citep{Euclid}. The key technical innovation of this work lies in the use of gradient-boosting algorithms that can capture complex nonlinear relations in the data \citep[e.g.][]{comparison_supervised}. This makes them well suited for the task of preselecting LAEs.\par

This paper is organized as follows. Section \ref{chapter_data} describes the data we used. Section \ref{chapter_ml} outlines the algorithmic framework and details the methods with which we evaluated the model, divided the data, and optimized the hyperparameters. In Section \ref{chapter_identification} we present the results of the machine-learning models. This includes an in-depth analysis of the training performance, predictions, and robustness to observational uncertainty perturbations. In Section \ref{section_discussion} we discuss the main results and also discuss the potential applications of this work to other surveys. Finally, Section \ref{chapter_conclusion} offers a summary of the key findings.\par

We adopt the same $\Lambda$CDM cosmological model as \citet{Cosmos2020} and \citet{sc4k} (H$_0$ = 70 kms$^{\text{-}1}$ Mpc$^{\text{-}1}$, Ω$_M$ = 0.3, and Ω$_\Lambda$ = 0.7). The magnitudes are given in the AB system \citep{ab}.

\section{Data} \label{chapter_data}

\hspace{0.3cm} We required a statistically significant sample of sources from LAE and non-LAE (nLAE) populations, where the latter represents other typical SFGs. They are both necessary for training a classifier to distinguish between them based on their broadband photometric properties. To achieve this, we used the SC4K \citep{sc4k} sample together with the COSMOS2020 catalog \citep{Cosmos2020}.\par

\subsection{SC4K}

\hspace{0.3cm} SC4K \citep{sc4k} is a survey in the COSMOS field \citep{cosmos_2007,Cosmos_Survey} that is designed to identify LAEs at $z \in [2,6]$. It uses 12 medium and 4 NBs over a 2 deg$^2$ area, covering a comoving volume of $\sim 10^8$ Mpc$^3$. This sample includes 3,908 sources that were selected using an observed EW threshold of EW $>50 \times (1+z)\AA$ and $\Sigma >3$, where the latter is the emission-line or excess significance \citep[e.g.][]{bunker_1995}, which measures the degree to which the observed counts in a broadband filter deviate from the expected counts based on measurements in a NB filter. The typical EW threshold is $25\AA$ \citep[e.g.][]{santos_2016}. With almost twice this value, contamination by lower redshift line-emitters is less likely. We also grouped these sources (according to the corresponding medium-band filter) into specific redshifts following Table 3 in \citet{sc4k}. For each selection filter, we attributed the average value of the redshift interval (see \citet{sc4k} for a detailed description of the sample selection.)\par

\subsection{COSMOS2020}

\hspace{0.3cm} In addition to LAEs, we required a sample of nLAE sources (i.e., sources that are present in COSMOS2020 but not in SC4K) within the same field and redshift range to ensure a well-balanced dataset for comparative studies. For this purpose, we used the COSMOS2020 CLASSIC catalog \citep{Cosmos2020}, which is based on the COSMOS survey \citep{Cosmos_Survey}. We did not use the FARMER catalog because we required reliable photometry, which is primarily restricted to the UltraVISTA \citep{ultravista} footprint as a result of mask requirements. Although it has fewer sources, CLASSIC ensures a complete and uniform spatial coverage for the analysis (i.e., for a comparison with the SC4K; see \cite{Cosmos2020}).\par

\begin{figure*}
\centering
\includegraphics[scale=0.33]{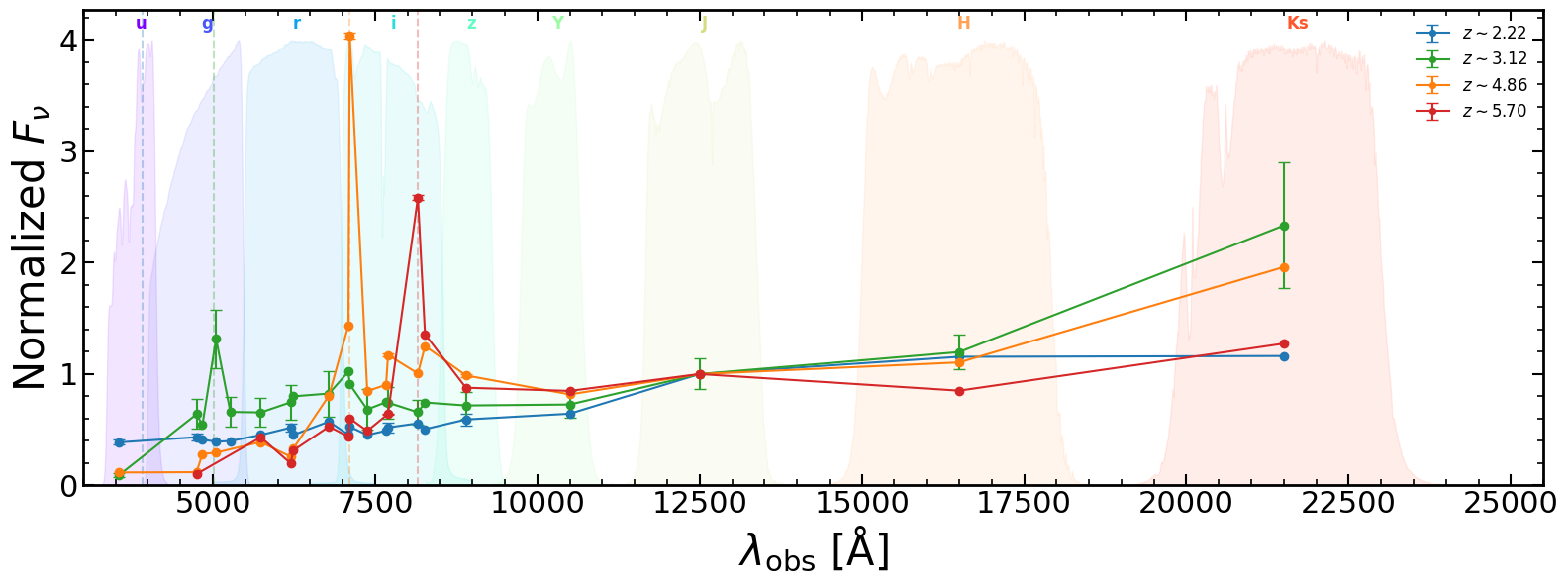}
\caption{Median SEDs of LAEs at different redshifts and normalized filter profiles of the broad bands. We  also include the broadband transmission curve as background shading for context, and the vertical line in the NB central wavelength shows the expected location of Ly$\alpha$.}
\label{LAE_templates}
\end{figure*}

We started by restricting the photometric redshift range ($2 < z < 6$) of the COSMOS2020 sample to match the SC4K survey because we lack comparable LAE samples outside this range to construct a meaningful classifier. The colors and intrinsic fluxes are highly dependent on redshift. We also excluded the SC4K sources from the COSMOS2020 sample. We cross-matched the two catalogs using TOPCAT \citep{topcat} (with a maximum matching error of 1'') to integrate photometric information into SC4K. Following these steps, COSMOS2020 and SC4K (hereafter LAE sample) contained 196,713 and 3,346 sources, respectively.\par

We assumed that galaxies in the COSMOS2020 catalog at 2<z<6 that were not detected in SC4K are not LAEs. We expected some LAE contaminants from COSMOS2020, however, in particular, sources with Ly$\alpha$ emission with low EWs, because SC4K relies on human validation, which is prone to failure. It also only covers certain redshift slices, possibly missing LAEs outside of these intervals. To mitigate this issue, and because LAEs are a subdominant population \citep[e.g.][]{cassata_2015}, we selected five different subsets and did not rely on a single sample. This was to ensure that first, the number of nLAEs extracted from COSMOS2020 was equal to the size of the LAE sample to exclude imbalances in the sample that might bias the model toward the majority class \citep[e.g.][]{bias_majority}. We also ensured that nLAE sources were drawn from the COSMOS2020 sample, following the same i-band magnitude and redshift distribution as the LAE sources. Specifically, the i-band magnitude ranges from 18 to 34 (minimum and maximum values for the i-band magnitude in the LAE sample, respectively), divided into intervals of one magnitude each. The redshift ranges from 2 to 6, also divided into intervals of one unit. This ensured that any potential differences are not dominated by distributions in redshift or UV brightness from our choice of samples. Following these steps, we obtained five subsamples that each contained the same LAE sources and different nLAE sources, extracted as described above.\par

The magnitude distributions in the two samples show (Figure \ref{mag_distribution}) that the LAE sample in the g band contains brighter sources than the nLAE sample. This highlights the possible bias toward brighter LAEs, which stems from that fact that the LAE sample was obtained using NB filters. On the other hand, no significant differences were found between the populations in the other bands. With the exception of the $g$, $r$, $i$, and $z$ bands, there is a significant number of missing values (e.g., NaN) in both samples, which can either signify nondetections or no image coverage. A nondetection is often essential for selecting high-redshift objects (e.g., Lyman-break selections). Moreover, the gradient boosting algorithms we used can effectively handle missing values, and these consequently pose no problem for our work.\par

\section{Machine-learning framework} \label{chapter_ml}

\hspace{0.3cm} In light of the accelerated growth in size and complexity of astronomical datasets, astronomers are developing automated tools to detect, characterize, and classify objects using these rich and complex datasets \citep[e.g.][]{ReviewMLAstro,mlastro_4, ML_Astro_2, ML_Astro_3}. Our focus here is supervised learning to train models that can accurately classify LAEs based on broadband photometric labels.\par

\begin{figure*}
\centering
\includegraphics[scale=0.35]{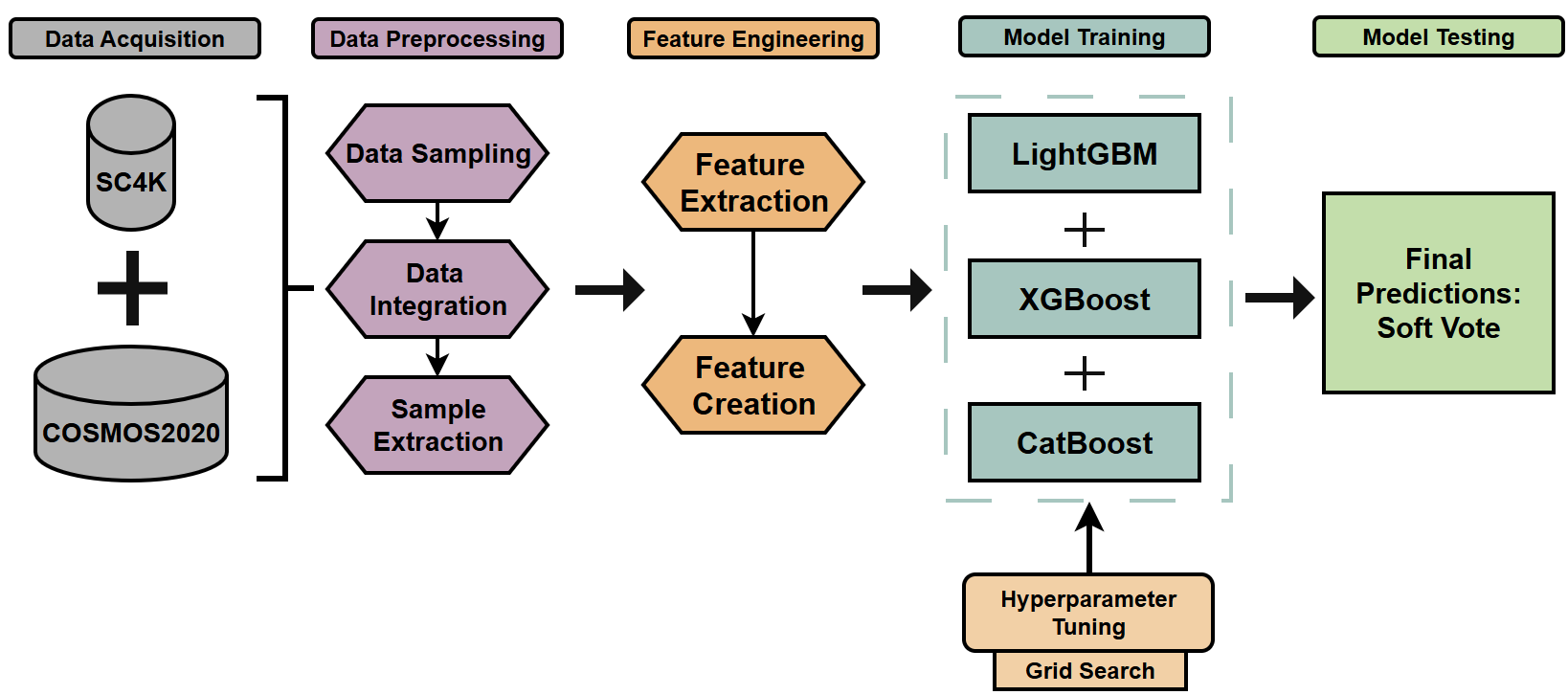}
\caption{ML flowchart of our framework. It begins with the acquisition of the publicly available SC4K and COSMOS2020 catalogs. The acquired data are then preprocessed, which involves sampling, integrating, and extracting the samples to be used by the ML models. This is followed by extraction of fluxes and magnitudes as features together with the creation of colors using the latter. The training phase involves the optimization of hyperparameters using  a grid search for each algorithm: LightGBM, XGBoost, and CatBoost. Finally, the final target label is obtained by combining all the individual predictions into a single prediction using soft voting.}
\label{flowchart}
\end{figure*}

\subsection{Feature engineering}

\hspace{0.3cm} The selected features for our work consist of broadband photometric properties: fluxes, magnitudes, and colors in the optical and near-infrared (NIR) spectra. The nine bands and their corresponding instruments, central wavelength, and observed depth are listed in Table \ref{bands}. For each band, we directly extracted fluxes and magnitudes with aperture magnitude APER3 from COSMOS2020, and we calculated the unique permutations of the broadband colors using the magnitudes. This resulted in 54 features. We worked with the optical and NIR properties because they help us in our overarching goal of creating a model that is applicable to other large surveys that are limited to these wavelengths.\par

The data-splitting strategy we followed involved dividing the data into $75\%$ for the training set, $10\%$ for the validation, and $15\%$ for the testing \citep[following the division suggested in][]{data_splitting}. This also ensured a stratified and shuffled division.

\begin{table}[ht]
\caption{Photometric bands and the corresponding instrument, central wavelength, and observed depth ($3''$).}
\centering
\begin{tabular}{cccc}
\toprule
Band & Instrument/Telescope & $\lambda$ $[Å]$ & Depth [mag] \\
\midrule
$u^*$ & MegaCam/CFHT & 3858 & 27.1 \\
$g$ & HSC/Subaru & 4847 & 27.5 \\
$r$ & HSC/Subaru & 6219 & 27.2 \\
$i$ & HSC/Subaru & 7699 & 27.0 \\
$z$ & HSC/Subaru & 8894 & 26.6 \\
$Y$ & VIRCAM/VISTA & 10216 & 26.1 \\
$J$ & VIRCAM/VISTA & 12525 & 25.9 \\
$H$ & VIRCAM/VISTA & 16466 & 25.5 \\
$K$ & VIRCAM/VISTA & 21557 & 25.2 \\
\bottomrule
\end{tabular}\relax
\tablefoot{The instruments we used are the Canada-France-Hawaii telescope (CFHT) and MegaCam instrument \citep{megacam}, the Hyper Suprime-Cam (HSC) \citep{miyazaki_2018}, and the VIRCAM instrument on the VISTA telescope from the UltraVISTA survey \citep{ultravista}.}
\label{bands}
\end{table}

Figure \ref{LAE_templates} shows the median spectral energy distributions (SEDs) of LAEs at different redshifts, normalized to the J band, overlaid with the normalized transmission curves of the broadband filters we used. For each NB redshift bin in COSMOS2020 (NB392, NB501, NB711, and NB816), we selected galaxies with at least six photometric bands with positive flux. For each filter, the median flux across these galaxies was computed based on the positive flux values, and the uncertainty was estimated using the median absolute deviation divided by the square root of the number of galaxies that contribute to that filter (NB392: 100 valid sources, NB501: 11 valid sources, NB711: 42 valid sources, and NB816: 108 valid sources). The median SED was then normalised by the flux in a reference band redward of Ly$\alpha$ (J band) for a direct comparison of the shape and emission excess (i.e., Ly$\alpha$ bump) in the bins.

\subsection{Algorithms}

\hspace{0.3cm} We used ensemble learning \citep[e.g.][]{ensemble,pc_2022,humphrey_2023,pc_2024}, which is an approach in machine learning that seeks a better predictive performance by combining the outputs of several models. In contrast to ordinary learning approaches that try to construct one learner from training data, ensemble uses a set of models and then aggregates their predictions in order to achieve a stronger predictive model with a better accuracy and an increased robustness \citep[e.g.][]{ensemble_2}. Within ensemble learning, we used gradient-boosting algorithms \citep{Friedman_2001, gbm_explained}, where the base learners are generated sequentially and thus explore the dependence between the base learners. This dependence can be used to good advantage as the performance can be boosted by minimizing the errors made by the previous model. This improves the overall performance.\par

We implemented three widely used current \citep{state_art_gbm} gradient-boosting algorithms: CatBoost \citep[version 1.2.3;][]{catboost}, XGBoost \citep[version 4.3.0;][]{xgboost}, and LightGBM \citep[4.3.0;][]{lgbm}. These models have been widely used for industry and research problems \citep[e.g.][]{gbm_application_1,gbm_application_2,gbm_application_3}. They all belong to the family of gradient-boosting decision trees, which build an ensemble of decision trees in a sequential manner. Each new tree is trained to correct the prediction errors made by the ensemble so far, using gradients of the loss function \citep[e.g.][]{gbm_explained}. The key idea behind boosting is that while each individual tree may be a weak learner, their collective combination results in a strong and flexible model.\par

Despite this shared foundation, the three frameworks differ in several key structural aspects. CatBoost creates symmetric trees with the same splitting rules at each level, which helps with speed and reduces overfitting. In contrast, LightGBM and XGBoost build asymmetric trees that can have different split conditions at the same depth. When it grows trees, LightGBM uses a leaf-wise method that focuses on the branches with the highest potential to reduce error. This makes the models smaller and faster, but increases the risk of overfitting. CatBoost and XGBoost use a level-wise approach and grow all branches evenly. Regarding splitting methods, CatBoost checks all feature-split options for each leaf and selects the option with the least penalty. LightGBM speeds the computation up by focusing on data points with the largest errors. XGBoost uses a slower histogram-based method to determine the best splits. These differences mean that a combination of the three algorithms through soft voting takes advantage of their complementary strengths and helps us to reduce individual model biases.\par

To determine the best combination of the hyperparameters, we used Grid Search \citep{hyper_opt}, which systematically explores a predefined set of values and selects the combination that optimizes a chosen performance metric. Not all the hyperparameters have the same effect on the performance variation of the algorithms \citep[e.g.][]{hyperparameter_opt_2,hyperparameter_opt_1}. We therefore followed the suggestions from the documentation of the three algorithms\footnote{\href{https://lightgbm.readthedocs.io/en/stable/}{LightGBM},  
\href{https://xgboost.readthedocs.io/}{XGBoost},  
\href{https://catboost.ai/docs/}{CatBoost}}  for guidance about the most effective hyperparameters for each. All the hyperparameters we optimized are listed in Appendix \ref{appendix_tune}. The summary of the optimization procedure is shown in Table \ref{table:optimization_grid}, and the default hyperparameters and grids we used are described for each algorithm. Table \ref{final_hyper-parameter_set} lists the final set of hyperparameters that was obtained using Grid Search for each algorithm and sample.\par
Figure \ref{flowchart} shows a flow diagram that contains an overview of the whole framework. 

\subsection{Evaluation metrics}

\hspace{0.3cm} We wished to evaluate the predictive ability, generalization capability, and overall quality of our models. In this subsection, we briefly introduce the evaluation metrics that we used mainly. The scores for the metrics we present below varied between zero and one, where one is the maximum score.\par

Precision (also called purity) is the fraction of relevant instances (known as true positives) among the retrieved instances,
\begin{equation}
    \text{Precision} = \frac{TP}{TP+FP},
\end{equation}
where TP is the number of true positives in a given class, and FP is the number for false positives in a given class.
Recall, also known as completeness, is the fraction of relevant instances that were retrieved,
\begin{equation}
    \text{Recall} = \frac{TP}{TP+FN},
\end{equation}
where FN is the number of false negatives in a given class.\par

\begin{table*}
\small
\caption{Average validation and test-set classification evaluation metrics for LightGBM, CatBoost, and XGBoost without hyperparameter optimization, that is, with the default settings. The standard deviation of each metric is also included.}
\centering
\begin{tabular}{lcccc|cccc}
\toprule
& \multicolumn{4}{c}{Validation} & \multicolumn{4}{c}{Test}\\
\cmidrule(lr){2-5} \cmidrule(lr){6-9}
Model & F1 & Acc & Prec & Rec & F1 & Acc & Prec & Rec \\
\midrule
LightGBM & 0.859$\pm$ 0.011 & 0.860$\pm$ 0.012 & 0.869$\pm$ 0.021 & 0.848$\pm$ 0.010 & 0.871$\pm$ 0.009 & 0.873$\pm$ 0.008 & 0.887$\pm$ 0.012 & 0.855$\pm$ 0.012\\
XGBoost & 0.859$\pm$ 0.016 & 0.860$\pm$ 0.018 & 0.870$\pm$ 0.032 & 0.848$\pm$ 0.009 & 0.865$\pm$ 0.008 & 0.867$\pm$ 0.008 & 0.879$\pm$ 0.010 & 0.852$\pm$ 0.009\\
CatBoost & 0.864$\pm$ 0.008 & 0.865$\pm$ 0.009 & 0.872$\pm$ 0.017 & 0.857$\pm$ 0.004 & 0.873$\pm$ 0.006 & 0.876$\pm$ 0.006 & 0.890$\pm$ 0.008 & 0.857$\pm$ 0.005\\
\bottomrule
\end{tabular}
\label{tab:model_performance_sets}
\label{tab:default_metrics_classification}
\end{table*}

These two metrics are different, and it is therefore important to understand they are correctly interpreted and which importance should be assigned to each. Precision is more important when the cost of false positives is high, that is, in situations when misclassifying an instance as positive has serious consequences. On the other hand, recall is more effective when the cost of false negatives is high, that is, when the consequence of misclassifying an instance as negative is relevant. Because they are different and their associated meaning differs, it is useful to combine them using the F1-score, which is the harmonic mean of precision and recall,
\begin{equation}
    \text{F1}  = 2\times\frac{\text{Precision}\cdot \text{Recall}}{\text{Precision} + \text{Recall}}.
\end{equation}

The objective is to identify a substantial number of LAEs while minimizing false positives. The F1-score is therefore beneficial because it provides a balance between these two factors. Accuracy is another objective. This is the fraction of predictions that the model derived correctly. It is formally defined as
\begin{equation}
    \text{Accuracy} = \frac{TP+TN}{TP+TN+FN+FP},
\end{equation}
and it should be used preferably when the number of objects in each class is roughly the same. This is true for our samples.\par

We calculated these four metrics for the models we obtained because each one of them attributes different importances to TP, TN, FP, and FN. \par

\begin{table*}
\small
\caption{Average validation and test-set classification evaluation metrics for LightGBM, CatBoost, and XGBoost with hyperparameter optimization. The standard deviation of each metric is also included.}
\centering
\begin{tabular}{lcccc|cccc}
\toprule
& \multicolumn{4}{c}{Validation} & \multicolumn{4}{c}{Test}\\
\cmidrule(lr){2-5} \cmidrule(lr){6-9}
Model & F1 & Acc & Prec & Rec & F1 & Acc & Prec & Rec \\
\midrule
LightGBM & 0.875$\pm$ 0.012 & 0.876$\pm$ 0.013 & 0.888$\pm$ 0.024 & 0.862$\pm$ 0.006 & 0.870$\pm$ 0.012 & 0.872$\pm$ 0.012 & 0.889$\pm$ 0.015 & 0.851$\pm$ 0.011\\
XGBoost & 0.871$\pm$ 0.011 & 0.873$\pm$ 0.013 & 0.886$\pm$ 0.024 & 0.858$\pm$ 0.005 & 0.866$\pm$ 0.011 & 0.869$\pm$ 0.011 & 0.885$\pm$ 0.014 & 0.845$\pm$ 0.011\\
CatBoost & 0.866$\pm$ 0.015 & 0.868$\pm$ 0.013 & 0.878$\pm$ 0.015 & 0.856$\pm$ 0.030 & 0.863$\pm$ 0.012 & 0.866$\pm$ 0.010 & 0.882$\pm$ 0.008 & 0.845$\pm$ 0.022\\
\bottomrule
\end{tabular}
\label{tab:model_performance_sets}
\label{metrics_classification_tuned}
\end{table*}

\section{Results} 
\label{chapter_identification}

\hspace{0.3cm} In each one of the five samples presented in Section \ref{chapter_data} (containing the same number of LAEs and nLAEs), the three algorithms were trained and tested. This resulted in 15 different models. In the following section, we present the performance evaluation of the algorithms with and without hyperparameter optimization together with the application of these algorithms to the remainder of the COSMOS2020 sample at $z \in [2,6]$.

\subsection{Machine-learning training performance and predictions in COSMOS2020}

\hspace{0.3cm} Each algorithm was initially trained and evaluated using default hyperparameters with various input features: fluxes only, both fluxes and magnitudes, and a combination of fluxes, magnitudes, and colors. When we only used fluxes or fluxes and magnitudes, the averaged F1-scores (for five samples) ranged from $\sim 0.82-0.84$ in the test, depending on the algorithm. The inclusion of colors significantly enhanced the model performances and resulted in higher F1-scores of $\sim 0.86-0.88$. We found a clear advantage in using a more extensive set of features to improve the predictive accuracy of the models, even though the physical information in the fluxes and magnitudes was essentially the same. This result is consistent with the findings by \citet{pc_2022,humphrey_2023,pc_2024}. The results we present below were obtained with fluxes, magnitudes, and colors as features.\par

The classification evaluation metrics we computed for each model without the hyperparameter optimization are presented in Table \ref{tab:default_metrics_classification} by averaging the results of the five samples. CatBoost performed better than the other algorithms overall. The difference is not highly significant compared with LightGBM, but XGBoost slightly underperformed in comparison. The results of the three algorithms all have a lower recall ($\sim 85-86\%$) than the other three metrics. On the other hand, precision is clearly the highlighted metric ($\sim 88-89\%$). \par

We optimized the hyperparameters (described in Appendix \ref{appendix_tune}) for all models using LightGBM, XGBoost, and CatBoost. The results are summarized in Table \ref{metrics_classification_tuned}.  The validation set improved by $\sim 2-5\%$ when the hyperparameters were optimized compared to the default case. These improvements became marginal for the test set. In contrast to the default case, LightGBM slightly outperformed the others, especially in the test set. CatBoost performed more poorly in the test set than in the default case. In Appendix \ref{appendix_z_impact} we present more detailed results regarding the impact of redshift of the training and testing samples in the model performance. For this purpose, we show in Figure \ref{fig:cm_all_models} the average confusion matrices for each algorithm separated by redshift interval bins. The three algorithms tend to increase at $z>4$ in general for the number of false predictions. This effect is present in the testing and training samples, but the latter is less affected.

The models were applied to the COSMOS2020 dataset with a focus on sources within $z\in[2,6]$ while excluding the five samples we used for the training. This resulted in 179,469 previously unused sources (hereafter called the predictions sample). We evaluated how often these sources were predicted as LAEs in 15 models. Over $70\%$ of sources were not classified as LAEs by any model (52,707 were predicted by at least one model), while fewer than 5$\%$ (7,073 sources) were consistently predicted as LAEs by all 15 models. The average prediction scores (ranging from zero to one) in the 15 models was also analyzed by computing the mean of the scores assigned by each model. By default, a source was classified as an LAE when its mean score exceeded 0.5. Scores closer to zero corresponded to nLAEs. We verified that the prediction score of 21,440 sources was higher than 0.5, within which the score for 3627 LAE candidates was higher than 0.9.

\subsection{Validating predicted LAEs with spectroscopic information}

\label{validation}

\hspace{0.3cm} After predicting LAEs in the COSMOS field, it is relevant to validate some of these predictions using publicly available spectroscopic catalogs. This validation provides insights into the decision-making process of the model and helps us to confirm the nature of certain sources. For this purpose, we used the COSMOS spectroscopic redshift compilation\footnote{\href{https://github.com/cosmosastro/speczcompilation}{COSMOS Spectroscopic Redshift Compilation}} \citep{cosmos_compilation}. It contains redshifts for 97,929 unique objects from 108 different observing programs up to $z \sim 8$.\par

We cross-matched the predicted LAEs from our model with the entries in this compilation using TOPCAT \citep{topcat}, which applies the SKY algorithm with a maximum matching error of $1''$. We restricted the results for $z \in [2,6]$ and applied a quality flag $Q_f > 1$, which ensured a confidence level $\geq80\%$. We also excluded all surveys without public information or without wavelength coverage for a $Ly\alpha$ detection. A source was considered a match when it had a rounded average prediction score higher than $0.5$ and was confirmed in any of the surveys available in \citet{cosmos_compilation}. Following these steps, we found 7 LAE confirmations in the VIMOS Ultra-Deep Survey (VUDS) \citep{vuds_1,vuds_2}, 15 in the COSMOS Ly$\alpha$ Mapping And Tomography Observations (CLAMATO) Survey \citep{clamato_1,clamato_2}, and 8 in the Deep Imaging Multi-Object Spectrograph (DEIMOS) \citep{deimos} with a Ly$\alpha$ EW higher than $20\AA$. Additionally,  10 were found in \citet{gto_muse}, all identified as Ly$\alpha$ emitters by the authors and with a Ly$\alpha$ EW higher than $20\AA$. In the Hobby-Eberly Telescope Dark Energy Experiment (HETDEX) \citep{hetdex_1,hetdex_2,hetdex_3}, 15 were identified as LAEs. Finally, 3 sources were studied in \citet{smuvs}, 1 by \citet{ning_2020}, and 1 source was presented in \citet{candelsz7}. In total, we obtained the cross-match of 60 LAEs with public spectroscopic catalogs in the COSMOS field. This corroborates that our models are able to effectively identify LAE candidates. The list of all the sources is provided in Appendix \ref{spectroscopic_cosmos}, and three example spectra are shown in Figure \ref{validation_plot_example}.

\subsection{Assessing the robustness of the classification models} \label{classification_generalization}

\hspace{0.3cm} Instrument measurements always include a degree of uncertainty over the measured magnitudes/fluxes. We assessed the impact of the level of uncertainty in our models by retraining the model with a dataset in which the magnitudes/fluxes were perturbed by their observational uncertainties. In each training/test sample, every source was perturbed by creating a Gaussian distribution with a mean equal to the flux value and a sigma equal to the flux error. We then drew a random value out of this distribution. The perturbed fluxes were then converted into AB magnitudes, which in turn were used to calculate the colors. Previously, fluxes and magnitudes were extracted from COSMOS2020, but we now wished to propagate the pertubations from the fluxes into the magnitudes. This justifies the conversion. This was done for every source in the five samples 100 times each for every band. We thus created five perturbed samples that were then used to train and test new models that were then compared with the unperturbed models we presented and discussed above.\par

The perturbations can be separated into three different cases:\par

\begin{enumerate}
\item Perturbations in the input data. The new models were trained and tested in these new  perturbed samples. We therefore ought to investigate the effect of the perturbations on the model performance and the consequences when predicting in unseen data.
\item Perturbations in the predictions table. We also introduced these variations in the predictions sample and then applied the unperturbed models to these 100 perturbed predictions tables.
\item Joint perturbations. We joined the perturbations performed in steps 1 and 2, that is, we applied the perturbed models to the perturbed predictions tables.
\end{enumerate}

We found a decrease in the F1-score by $\sim 5\% $ when we introduced perturbations compared to the unperturbed case, but no variation between iterations. This decrease was expected as a result of the introduction of variance in the data, but the F1-score still remained above $0.81$ for all algorithms. We also investigated the variation in the number of predicted LAEs with a prediction score over $0.9$ in the perturbed models for the three different cases. The number of predicted LAEs for the input data (1) decreased by $\sim 20\%$, while the perturbations in the prediction data (2) led to an increase of $\sim 40\%$ in the number of sources with a score higher than $0.9$. When the two perturbations were joined (3), the number of predicted LAEs was similar to the unperturbed case. The significant increase in the number of predicted LAEs when the prediction data were perturbed indicates that there might be a shift toward higher values of prediction scores in this case. This can also be affect by the fact that the prediction table is larger by almost two orders of magnitude than the input data. This leads to a higher chance of generating a LAE classification.\par

In order to further investigate this possible shift, we show in Figure \ref{perturbated_vs_unperturbed_classification} two-dimensional histograms of the perturbed and unperturbed prediction scores that we computed for the three perturbation cases. For the first perturbation case (1), the data points seem to follow a 1:1 relation with some scatter around this line, but there is some trend toward lower perturbed scores for sources with high unperturbed scores. The results clearly start to deviate from the 1:1 relation when the predictions table is perturbed (2), where in addition to the trend at the lower right side of the plot, there is also a clear shift of sources with low unperturbed scores toward higher perturbed scores. In the last case, the excess diminishes, but the trend at the lower right side of the plot is stronger.\par

\begin{figure}[htp]
\centering
\includegraphics[width=.35\textwidth]{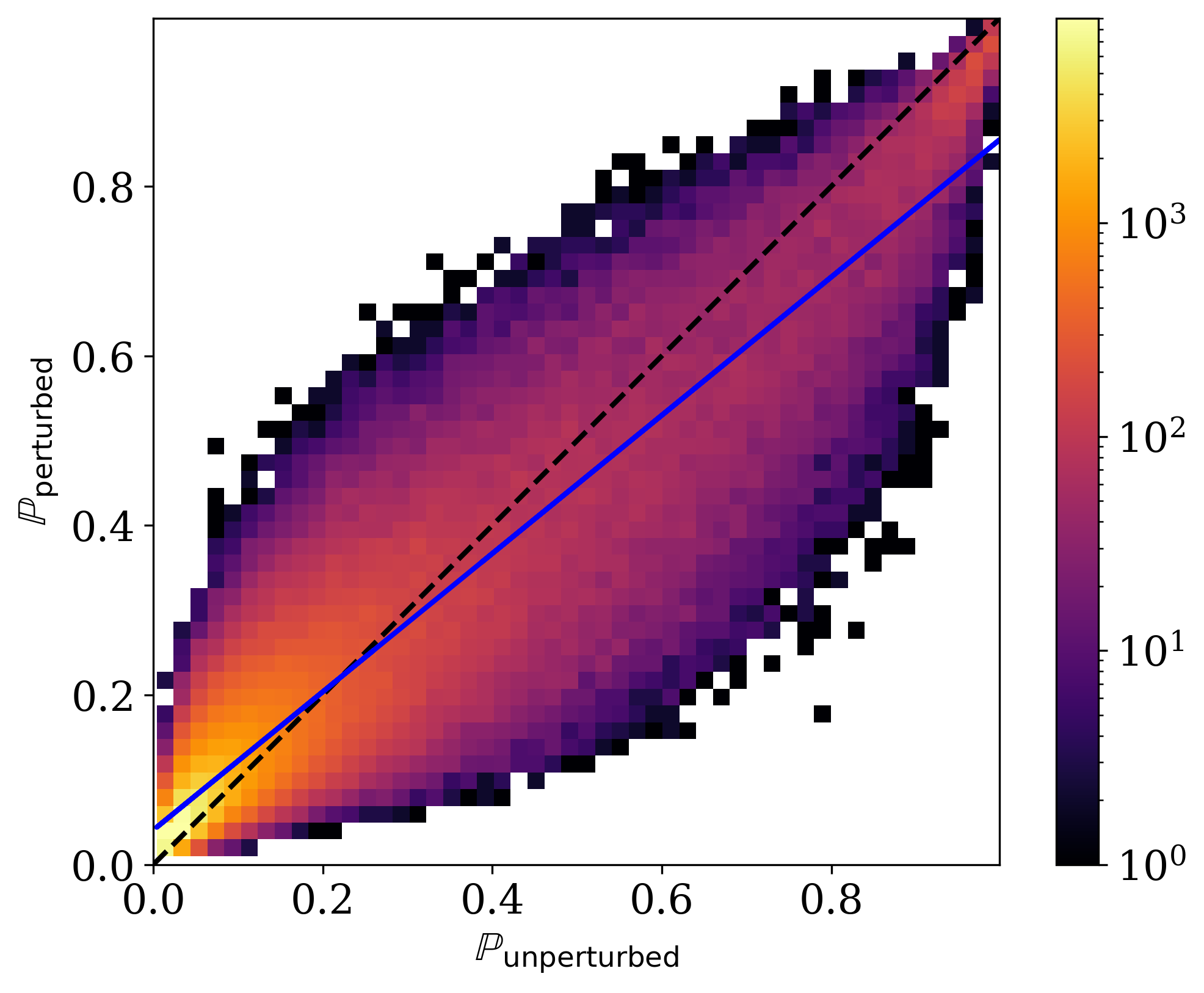}
\includegraphics[width=.35\textwidth]{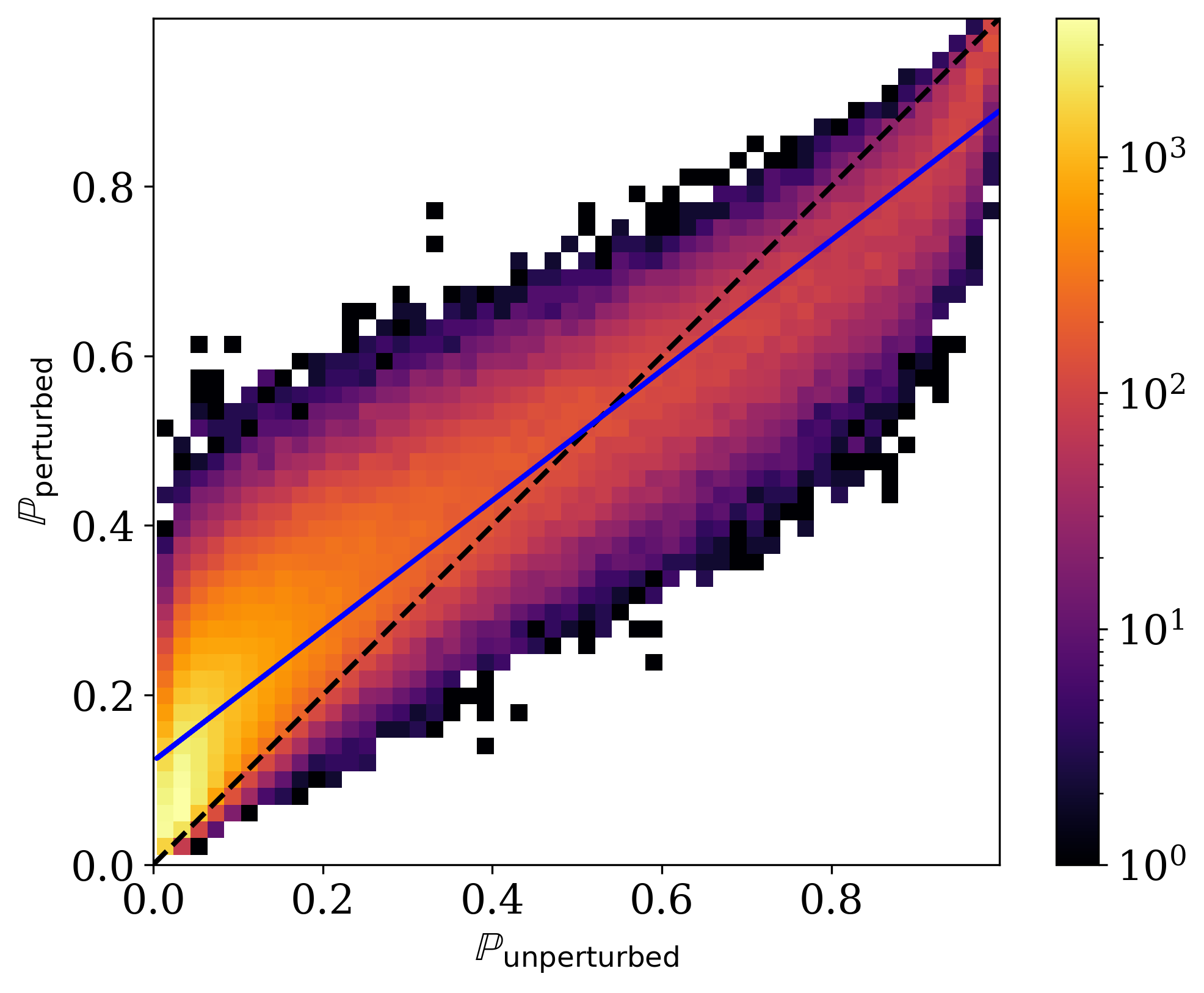}
\includegraphics[width=.35\textwidth]{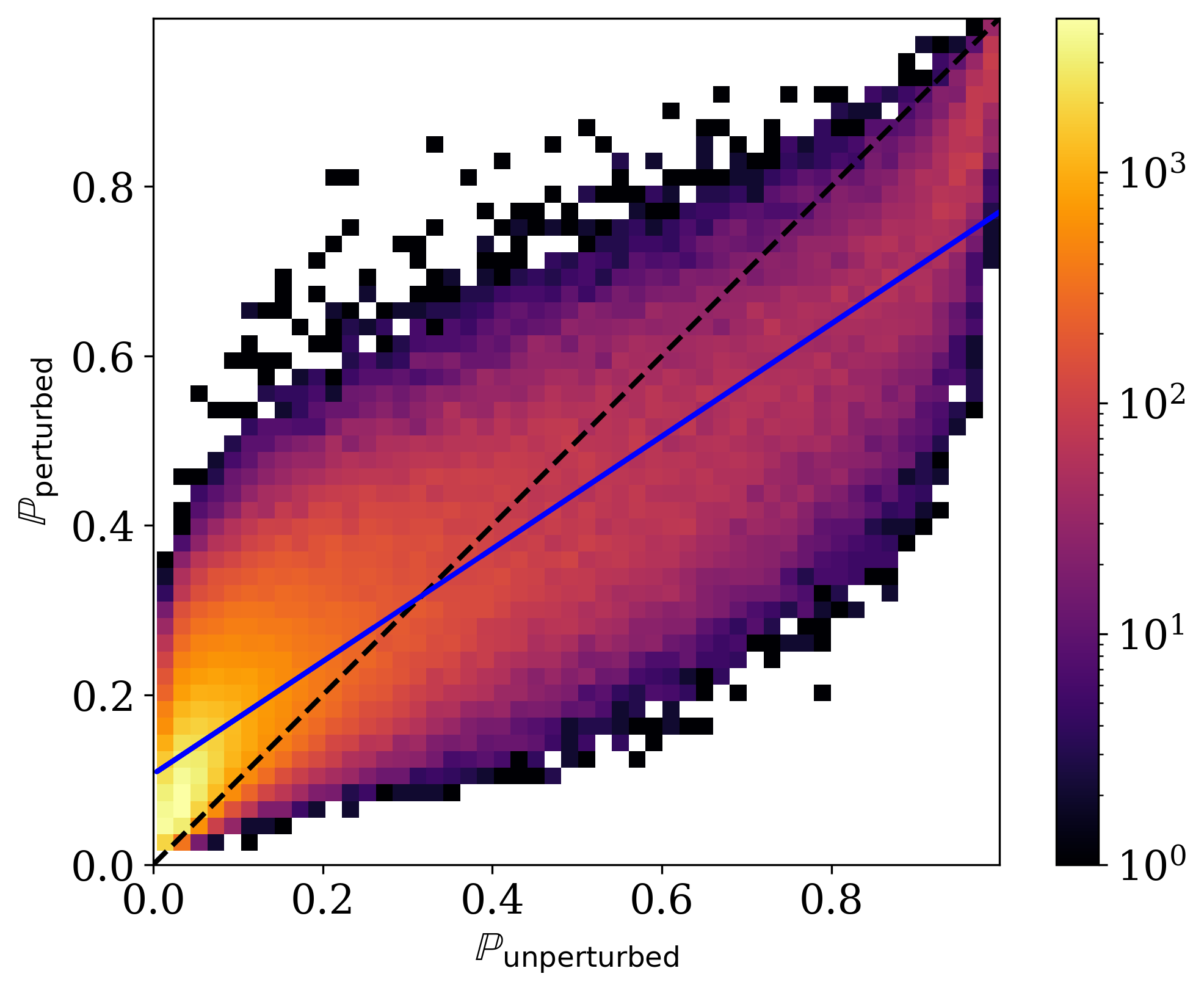}
\caption[Comparison of the perturbed and unperturbed prediction probabilities.]{Comparison of the perturbed prediction scores and unperturbed scores for the input data perturbation, prediction table perturbation, and joint perturbation cases in the panels from top to bottom, respectively. Average prediction scores over all the models were computed for each source. The blue lines represent a linear fit to the data points, and the dashed black lines show the 1:1 relation.}
\label{perturbated_vs_unperturbed_classification}
\end{figure}

\begin{figure*}[t!]
    \centering
    \begin{subfigure}{0.32\textwidth}
        \includegraphics[width=\linewidth]{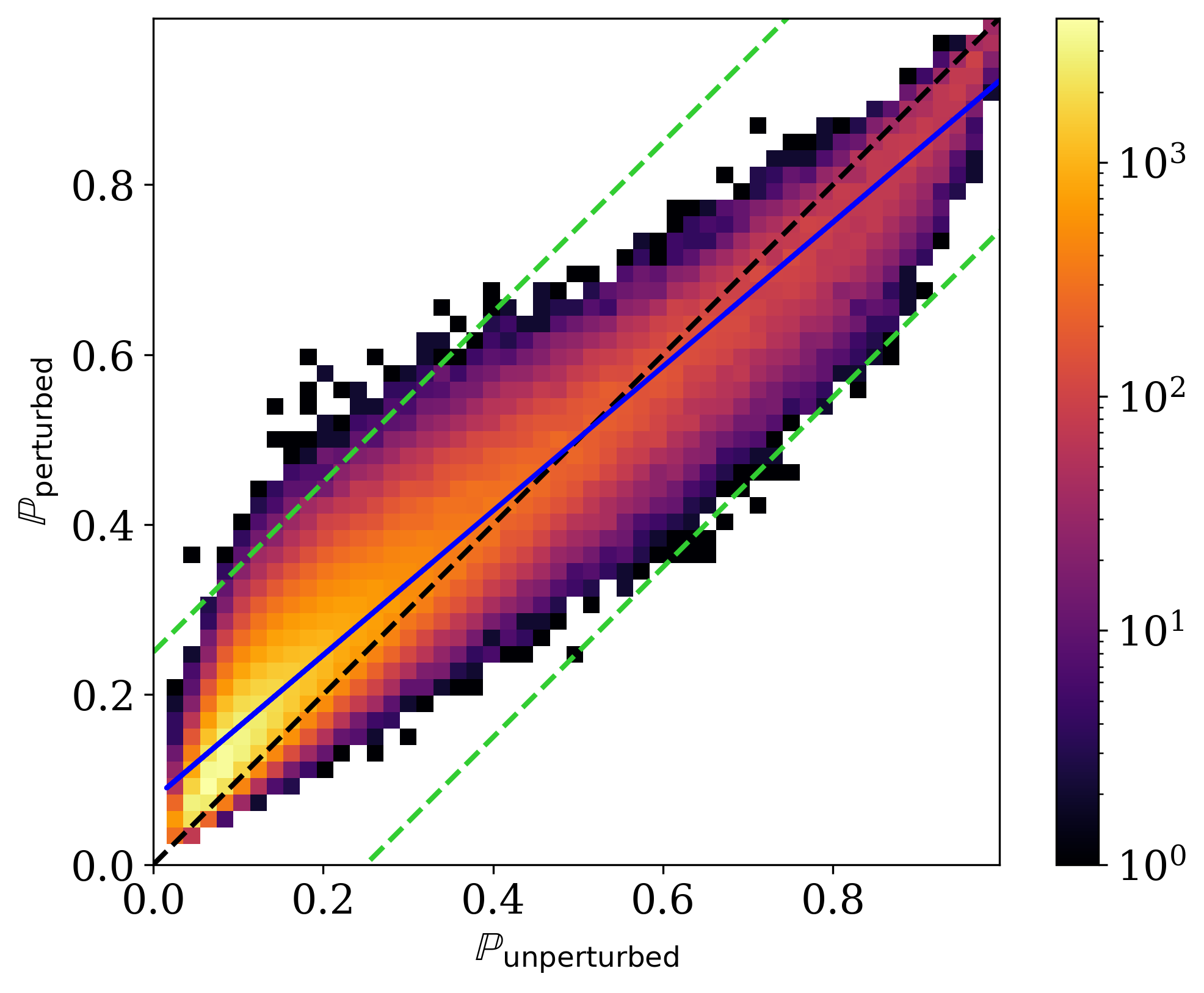}
        \caption{Fluxes}
    \end{subfigure}
    \hfill
    \begin{subfigure}{0.32\textwidth}
        \includegraphics[width=\linewidth]{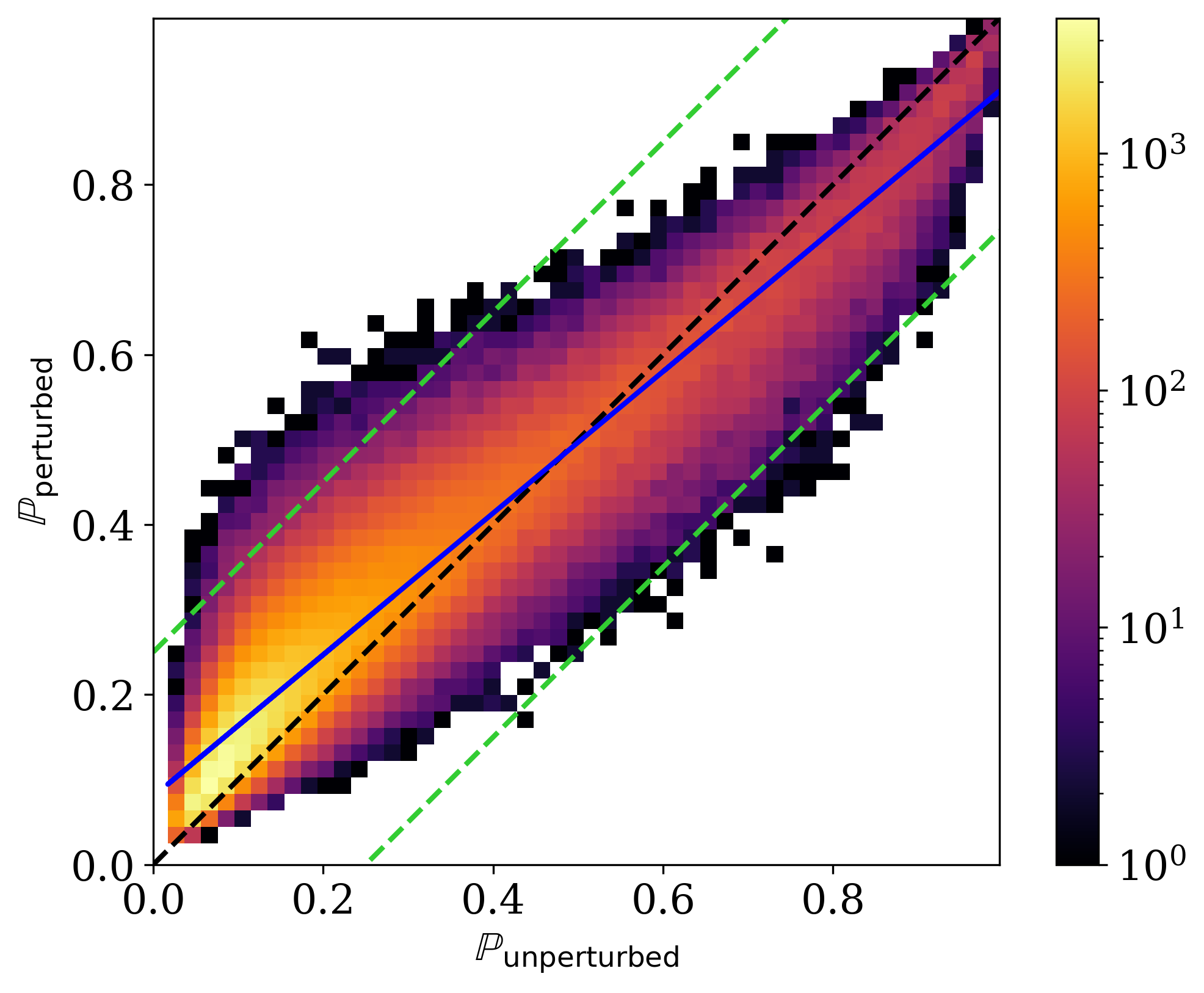}
        \caption{Fluxes and magnitudes}
    \end{subfigure}
    \hfill
    \begin{subfigure}{0.32\textwidth}
        \includegraphics[width=\linewidth]{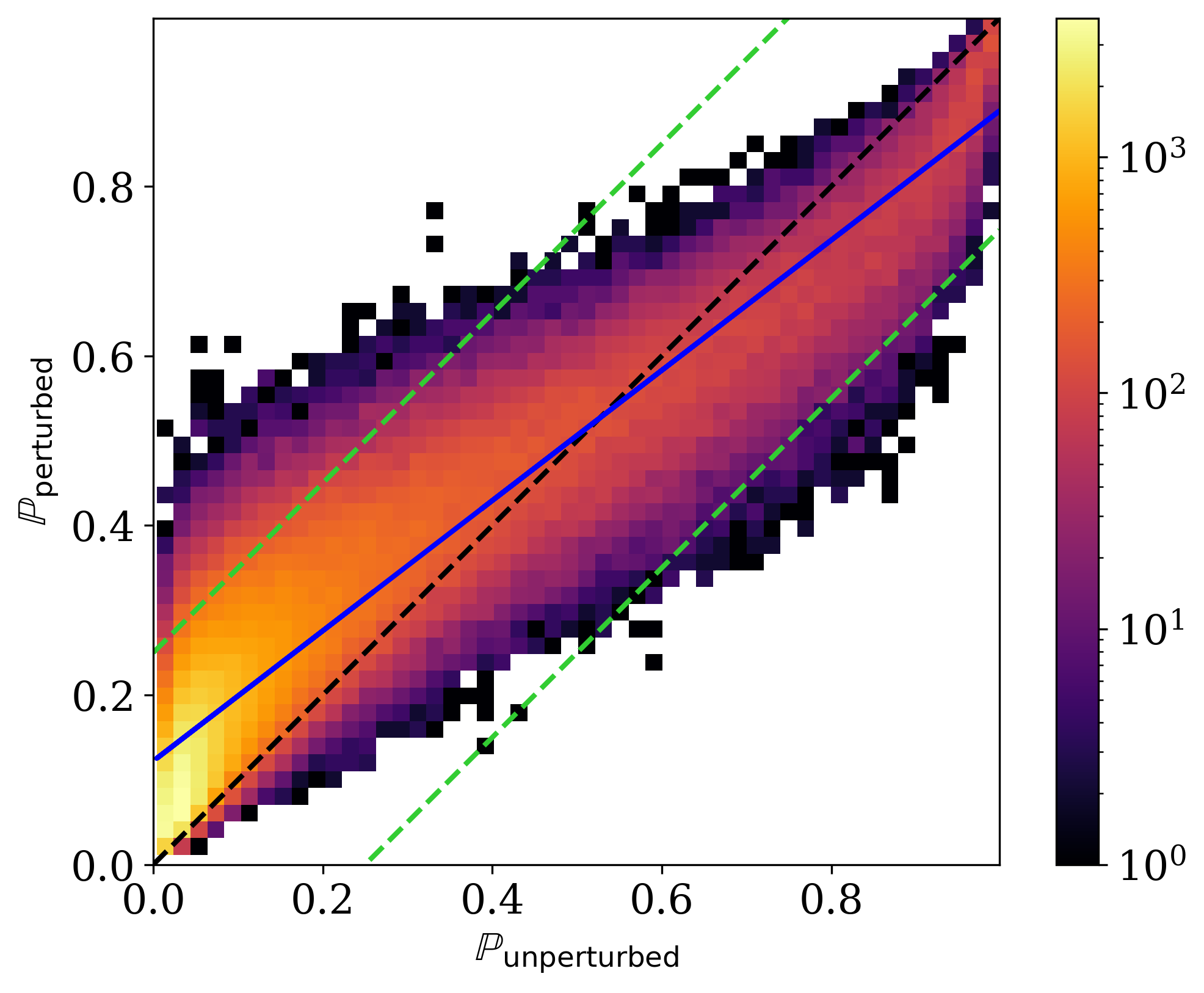}
        \caption{Fluxes, magnitudes, and colors}
    \end{subfigure}
    \caption[Comparison of the perturbations in the prediction data with different feature sets.]{Comparison of the perturbations in the prediction data results for fluxes (a), fluxes and magnitudes (b), and fluxes, magnitudes, and colors (c). The blue lines in each subplot represent a fit to the data, the dashed black lines show the 1:1 relation, and the dashed green lines show the 0.25 score deviation from the black line.}
    \label{perturbations_set_features}
\end{figure*}

To determine the cause of the stronger variation in the prediction data perturbations, we repeated the perturbations in the prediction data and separated them into three cases, depending on the feature set:} (i) fluxes, (ii) fluxes and magnitudes, and (iii) fluxes, magnitudes, and colors. The results are compiled in Figure \ref{perturbations_set_features}, separated for each case. When the fluxes alone were considered, the 0.25 range contains 237 sources. When magnitudes were added to the feature set, this number increases to 763. Finally, for the fluxes, magnitudes, and colors, the number of outliers increases massively to 5545. As the number of features increases, the scatter around the 1:1 relation between unperturbed and perturbed prediction scores therefore increases as well, which means that the bumps in the left and right parts of the plots become more pronounced. This is especially true when colors are introduced, which might be related to the double perturbation needed to compute each color. Colors imply two independent and random perturbations, one per band, whereas for fluxes/magnitudes, only one error/perturbation was considered.\par

\section{Discussion}
\label{section_discussion}

\hspace{0.3cm} The metrics for the default (Table \ref{tab:default_metrics_classification}) and optimized models (Table \ref{metrics_classification_tuned}) show robust models that are able to identify LAE candidates based on broadband photometry data in the optical and NIR alone. The small improvements with hyperparameter optimization are consistent with reports inother works \citep[e.g.][]{hyperparameter_tuning_improv_2,hyperparameter_tuning_improv_1}, which might indicate that the default algorithms are already near the maximum performance for these algorithms and these data. \par

More than 50,000 sources were predicted as LAE candidates in COSMOS2020 by at least one model, and approximately 7,000 of these were predicted by all 15 models we trained. Even when we only consider the latter, this is a highly significant number of predicted sources. If it were confirmed, it might more than double the existing SC4K sample. This confirms that previously unknown LAEs candidates might be detected in this way. We spectroscopically validated 60 LAEs in the COSMOS field (Section \ref{validation}), which again highlights that the models fulfill their potential in complementing traditional LAE detection methods. A possible explanation for the failure to detect these sources using NB and intermediate-band (IB) surveys might be that NB/IB filters only cover limited redshift ranges. They thus risk losing many sources between the gaps left at $z \in [2,6]$, which can now be explored (see Paulino-Afonso et al. in prep) because the whole redshift range can now be searched with the help of these models.\par

\subsection{Effect of the redshift on the model performance}

\begin{figure}[h]
\centering
\includegraphics[width=0.5\textwidth]{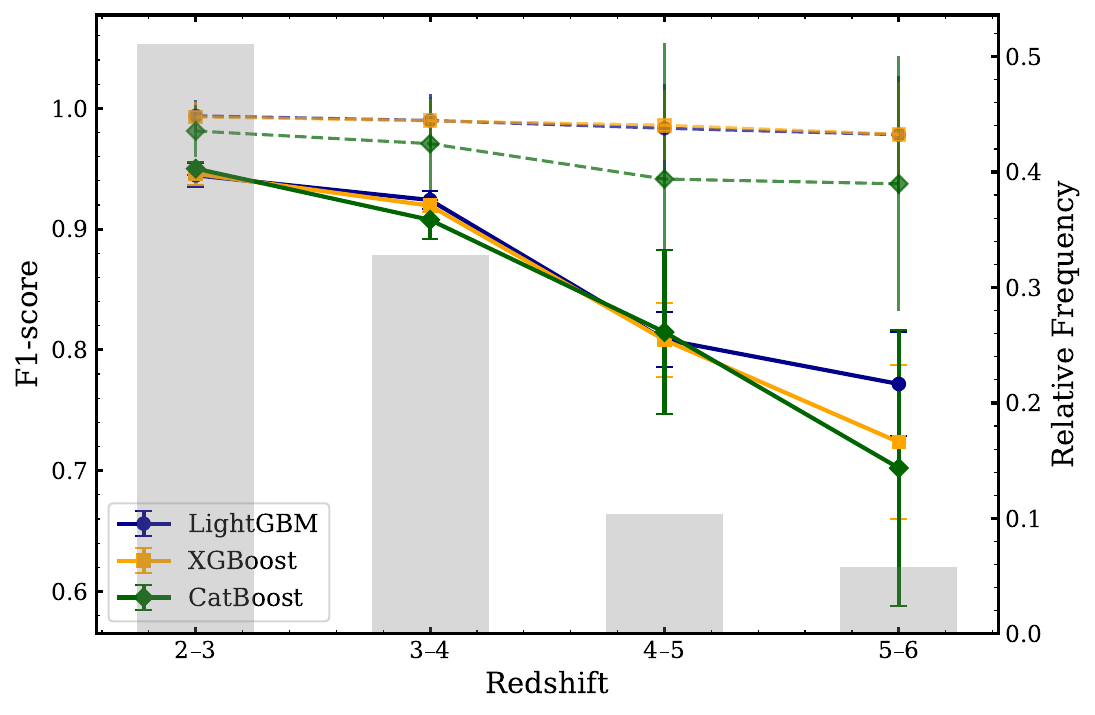}
\caption{Variation in the test and train (dashed) F1-scores as a function of redshift for the three gradient-boosting models LightGBM, XGBoost, and CatBoost. The overlaid histogram represents the normalized relative frequency of sources within each redshift bin in the training/testing samples. This combined view illustrates the relation between the model performance and data availability.}
\label{Redshift_impact_performance}
\end{figure}

\paragraph{} Figure \ref{Redshift_impact_performance} presents the test and train F1 scores as a function of redshift for the three algorithms. The figure shows that all three models performed strongly in the lower redshift bins (approximately $z \in [2,4]$), with F1 scores ranging from approximately 0.9 to 1.0. With increasing redshift ($z > 4$), however, the F1 scores consistently decrease in all models. This indicates a general degradation in the predictive performance. The overplotted histogram further confirms that the distribution of training and testing data is strongly skewed toward low-redshift sources, with a marked drop in the sample availability beyond $z \sim 4$.

This relation underscores a common challenge in machine learning: performance degradation in underrepresented regimes. The limited availability of high-redshift sources hampers the model performance when the data are scarce. The nearly parallel decline of the F1-scores in all three models suggests that architectural or optimization differences alone are insufficient to counteract the effects of this data imbalance.

To mitigate the performance drop at high redshifts, several strategies might be considered in principle. First, physically motivated data augmentation using the spectral synthesis code \texttt{CLOUDY} \citep{cloudy_2023} might be employed, for example. This method enables the generation of synthetic spectra based on astrophysical parameters, which can be redshifted and transformed into model-compatible features. While this approach introduces valuable data where observations are limited, it is limited, including potential mismatch with real observational noise, reliance on uncertain priors, and high computational cost. Nonetheless, when it is carefully applied, it might improve the model performance in data-sparse regimes.

An alternative to addressing the redshift imbalance is reweighting the training samples based on the inverse frequency of their redshift bin. This gives more weight to underrepresented high-redshift data and would improve the model performance in sparse regions without creating synthetic data. This method assumes that all redshift bins are equally informative and can cause overfitting or instability if high-weighted samples are noisy, however,. Additionally, a weight estimation can be unreliable in high-dimensional or limited data settings. Despite these challenges, reweighting can enhance the robustness across redshift intervals.

\subsection{Potential applicability in the era of large surveys} \label{classification_generalization}

\begin{table*}[ht]
\small
\caption{Average evaluation metrics in the test set compared with using a filter set adapted according to the wavelength coverage available in other surveys: Euclid, LSST, DES, and Roman. The standard deviations for each metric are also included.}
\centering
\begin{tabular}{lccccc}
\toprule
& \multicolumn{5}{c}{Test set} \\
\cmidrule(lr){2-6}
Survey & Algorithm & F1-Score & Accuracy & Precision & Recall \\
\midrule
{All filters} & LightGBM & 0.870$\pm$0.012 & 0.872$\pm$0.012 & 0.889$\pm$0.015 & 0.851$\pm$0.011 \\
& XGBoost & 0.866$\pm$0.011 & 0.869$\pm$0.011 & 0.885$\pm$0.014 & 0.845$\pm$0.011 \\
& CatBoost & 0.863$\pm$0.012 & 0.866$\pm$0.010 & 0.882$\pm$0.008 & 0.845$\pm$0.022 \\
\addlinespace
{Euclid} & LightGBM & 0.714$\pm$0.014 & 0.722$\pm$0.013 & 0.735$\pm$0.014 & 0.693$\pm$0.016 \\
& XGBoost & 0.714$\pm$0.016 & 0.724$\pm$0.019 & 0.740$\pm$0.025 & 0.690$\pm$0.012 \\
& CatBoost & 0.719$\pm$0.010 & 0.728$\pm$0.015 & 0.745$\pm$0.026 & 0.694$\pm$0.005 \\
\addlinespace
{LSST} & LightGBM & 0.864$\pm$0.008 & 0.867$\pm$0.008 & 0.887$\pm$0.009 & 0.842$\pm$0.010 \\
& XGBoost & 0.866$\pm$0.006 & 0.869$\pm$0.006 & 0.890$\pm$0.010 & 0.842$\pm$0.010 \\
& CatBoost & 0.861$\pm$0.008 & 0.865$\pm$0.006 & 0.887$\pm$0.008 & 0.838$\pm$0.018 \\
\addlinespace
{DES} & LightGBM & 0.797$\pm$0.013 & 0.802$\pm$0.014 & 0.819$\pm$0.021 & 0.777$\pm$0.007 \\
& XGBoost & 0.797$\pm$0.009 & 0.802$\pm$0.011 & 0.821$\pm$0.018 & 0.774$\pm$0.004 \\
& CatBoost & 0.800$\pm$0.007 & 0.806$\pm$0.007 & 0.825$\pm$0.014 & 0.777$\pm$0.014 \\
\addlinespace
{Roman} & LightGBM & 0.825$\pm$0.009 & 0.829$\pm$0.010 & 0.846$\pm$0.016 & 0.805$\pm$0.011 \\
& XGBoost & 0.827$\pm$0.008 & 0.831$\pm$0.008 & 0.847$\pm$0.010 & 0.809$\pm$0.010 \\
& CatBoost & 0.823$\pm$0.010 & 0.828$\pm$0.010 & 0.845$\pm$0.017 & 0.803$\pm$0.017 \\
\bottomrule
\end{tabular}
\label{generalization_metrics_classification}
\end{table*}

\hspace{0.3cm} One important goal of this work was to explore whether the models might be applied to other fields and surveys. This generalization is not straightforward as it involves conversion between different sets of filters, with different transmission curves. Our oversimplified approach aims to evaluate how these models perform when certain bands that we used are unavailable. As mentioned, our study used filters that observed COSMOS and covered wavelengths from 3560 Å to 23094 Å. Accordingly, we removed photometric bands from the feature set given to train the models and tried to mimic the filter wavelength coverage of potentially interesting surveys and missions to which this work might be applied. We provide a summary of each survey we considered for this exercise and the filter set we used to train the models in each case below.\par

\begin{itemize}
\item Euclid. The Euclid mission \citep{Euclid} will observe 15\addedAPA{,}000 $deg^2$ of extragalactic sky from the Sun–Earth Lagrange point L2. Among many other data, the Euclid plans include determining the photometry of at least one billion galaxies. To achieve this goal, the Euclid spacecraft has a visible-wavelength camera (the VISible instrument; VIS) and a near-infrared camera/spectrometer (the Near-Infrared Spectrometer and Photometer; NISP). The VIS instrument covers the wavelength between $5500 \text{–} 9000 \AA$, and NISP has three broadband filters (Y, J, and H) that cover the wavelength ranges of $9000 \text{–} 11920 \AA$, $11920 \text{–} 15440 \AA$, and $15440 \text{–} 20000 \AA$, respectively. To mimic the wavelength coverage of the VIS instrument, we considered the r, i, and z bands, and to mimic the coverage of the NISP instrument, we considered the Y, J, and H bands.

\item LSST. The LSST \citep{lsst} will cover about 18,000 $deg^2$ of the southern sky with six filters (u, g, r, i, z, and Y) in a wavelength interval from approximately $3500$ to $10600 \AA$. Consequently, only the homonymous bands available in the COSMOS2020 catalog were used, that is, the u$^*$, g, r, i, z, and Y bands.  
\item DES. The Dark Energy Survey (DES) \citep{des} astronomical survey is designed to study hundreds of millions of galaxies and to help us to understand the nature of dark energy. It uses the newly built Dark Energy Camera (DECam), which records images using five filters (g, r, i, z, and Y) that span wavelengths from $4000 \AA$ to $10800 \AA$. We therefore only used the g, r, i, z, and Y bands. 
\item Roman. The Nancy Grace Roman Space Telescope \citep{Roman} is a NASA infrared space telescope created to answer important questions in the areas of dark energy, exoplanets, and infrared astrophysics. It is equipped with a wide-field instrument (WFI) carrying eight filters spanning 4800 – 23000 Å. The covered wavelength range is large, and only the u$^*$ band was removed from our exercise.
\end{itemize}

After defining the changes in the features used to mimic the wavelength coverage of each survey presented above, we trained the models and ran them for each survey. The metrics we obtained for the test set are summarized in Table \ref{generalization_metrics_classification}.\par

We compared the results for Euclid and Roman, where the only differences were the removal of the $u^*$ and $g$ bands in Euclid and only the $u^*$ band in Roman. The effect from excluding the $u^*$ band is significant, as is reflected in the Roman results ($\sim 5\%$ decrease). This effect is even more pronounced in Euclid, where the additional removal of the $g$-band further amplified the difference significantly ($\sim 17\%$). These findings highlight the crucial role of these two bands in enhancing the performance of our models to accurately distinguish LAEs from non-LAEs. In contrast, the removal of the $J$, $H$, and $K$ bands had little effect on the performance, as evidenced by the LSST results ($<1\%$). This is also visible in DES results compared to Roman, where the difference is the inclusion of the $J$, $H$, and $K$ bands in the latter. This gained $\sim 2\%$ in the F1-Score value.\par

Our analysis emphasized the high importance of the $g$, and u$^*$ bands. When these bands were included, we were better able to measure the Lyman-break at $z\geq$ 3. This probably indicates that coverage of the Lyman-break in the photometry is important to assess the nature of these galaxies. Surveys that lack these bands will not be able to ensure a comparable classification performance. The NIR bands are less effective at distinguishing LAEs than optical bands, however. The reason might be that the $Y,J,H,K$ bands have a high fraction of missing data (Figure \ref{mag_distribution}), which reduces their contribution to the classification models. This means that the filter set is almost not affected at all when these bands are removed.\par

Through a feature importance analysis (Figure \ref{fp_ident}) of each algorithm, averaged over the five samples, we found that the g-r color is the most important feature in all the three algorithms. This color is a rough approximation of the $\beta$ slope \citep[e.g.][]{bowens_2009} at $z \sim 2-3$ (the bulk of the samples), which traces the attenuation by dust in the galaxy. At higher reshifts ($\geq3.5$), it starts to measure the Lyman break, which is an indirect measure of the redshift \citep[c.f.][]{steidel_1996} and the neutral gas content. If g-r is the most important feature, it likely means that dust is crucial for separating LAEs and nLAEs \citep{sergio_santos_2019}. This is consistent with dust being one of the primary materials that absorbs Ly$\alpha$ photons and prevents them from escaping from galaxies.

\section{Conclusion} \label{chapter_conclusion}

\hspace{0.3cm} We applied machine-learning techniques, specifically, the gradient-boosting algorithms LightGBM, XGBoost, and CatBoost, to identify LAEs candidates using broadband photometric data (fluxes, magnitudes, and colors) in the optical and NIR. Using SC4K and COSMOS2020, we extracted five samples with similar redshift and i-band magnitude distributions to ensure that we had comparable LAE and nLAE populations. We finally trained, tested, and analyzed the three algorithms in each one of the five samples, resulting in 15 models. We summarize our main conclusions below.

\begin{itemize}
    \item[-] The machine-learning models we developed demonstrated robust classification abilities with F1-scores from approximately $86\%$ to $88\%$ in the different algorithms. We also verified that our models presented a significant performance decrease at $z>4$, where training/testing data are scarce.
    \item[-] These models successfully identified over 7,000 LAEs candidates in unanimous agreement in all models, and over 50,000 were predicted by at least one model. This significantly increased the number of LAEs in the COSMOS field.
    \item[-] Validation with spectroscopic data from \citet{cosmos_compilation} confirmed 60 of our machine-learning predicted LAEs. This underscored the effectiveness of the approach.
    \item[-] The robustness of the models was tested against perturbations in data input by introducing the errors over the measured fluxes, which were then propagated to magnitudes and colors. This led to a small decrease of $5\%$  in the F1-score and showed that with more features, the scatter between the perturbed and unperturbed predictions increased.
    \item[-] We also found that removing the $u^*$ and $g$ bands highly affected the model performance (a decrease in the F1-score of $\sim 6-17\%$). This is problematic for surveys such as Euclid, LSST, and Roman. On the other hand, removing the most NIR bands (Y-,J-, and H-bands) had an effect of less than $1\%$. This shows that the LSST survey is well suited for the application of these models.
\end{itemize}

Throughout this work, we emphasized the capability of broadband photometric data, coupled with machine learning, to offer a cost-effective and scalable complement to traditional NB surveys and blind spectroscopy. This is extremely useful to process vast amounts of broadband photometric data quickly and to identify candidates over large sky areas. This is pivotal for upcoming data from large-scale surveys such as LSST and Euclid. Furthermore, we clearly understood that an immense number of additional LAEs can be identified beyond those already detected. This might expand the known population and contribute to a broader and more detailed map of the early Universe.

\section*{Data availability}

All the data used throughout this work is publicly available\footnote{\href{https://astroweaver.github.io/project/cosmos2020-galaxy-catalog/}{COSMOS2020},  
\href{https://cdsarc.u-strasbg.fr/viz-bin/cat/J/MNRAS/476/4725}{SC4K}}. The code needed to reproduce this work is also available on GitHub (\url{https://github.com/valeafonso/FLAEMING-Classification}), together with the ML models. The complete candidates table (with sources predicted by the 15 models) is only available in electronic form at the CDS via anonymous ftp to cdsarc.u-strasbg.fr (130.79.128.5) or via http://cdsweb.u-strasbg.fr/cgi-bin/qcat?J/A+A/.

\begin{acknowledgements}
We thank the anonymous referee for their valuable comments and suggestions, which have greatly improved this manuscript. This work was supported by Fundação para a Ciência e a Tecnologia (FCT) through the research grants UIDB/04434/2020 and UIDP/04434/2020, and the exploratory project EXPL/FIS-AST/1085/2021 (PI: Paulino-Afonso). APA acknowledges FCT support through the FCT Investigador FCT Contract No. 2020.03946.CEECIND. JF acknowledges FCT support through the Investigador FCT Contract No. 2020.02633.CEECIND/CP1631/CT0002. This work is based on observations collected at the European Southern Observatory under ESO programme ID 179.A-2005 and on data products produced by CALET and the Cambridge Astronomy Survey Unit on behalf of the UltraVISTA consortium. HETDEX is led by the University of Texas at Austin McDonald Observatory and Department of Astronomy with participation from the Ludwig-Maximilians-Universität München, Max-Planck-Institut für Extraterrestrische Physik (MPE), Leibniz-Institut für Astrophysik Potsdam (AIP), Texas A\&M University, Pennsylvania State University, Institut für Astrophysik Göttingen, The University of Oxford, Max-Planck-Institut für Astrophysik (MPA), The University of Tokyo and Missouri University of Science and Technology. Observations for HETDEX were obtained with the Hobby-Eberly Telescope (HET), which is a joint project of the University of Texas at Austin, the Pennsylvania State University, Ludwig-Maximilians-Universität München, and Georg-August-Universität Göttingen. The HET is named in honor of its principal benefactors, William P. Hobby and Robert E. Eberly. The Visible Integral-field Replicable Unit Spectrograph (VIRUS) was used for HETDEX observations. VIRUS is a joint project of the University of Texas at Austin, Leibniz-Institut für Astrophysik Potsdam (AIP), Texas A\&M University, Max-Planck-Institut für Extraterrestrische Physik (MPE), Ludwig-Maximilians-Universität München, Pennsylvania State University, Institut für Astrophysik Göttingen, University of Oxford, and the Max-Planck-Institut fur Astrophysik (MPA). The authors acknowledge the Texas Advanced Computing Center (TACC) at The University of Texas at Austin for providing high performance computing, visualization, and storage resources that have contributed to the research results reported within this paper. URL: http://www.tacc.utexas.edu Funding for HETDEX has been provided by the partner institutions, the National Science Foundation, the State of Texas, the US Air Force, and by generous support from private individuals and foundations. Based also on data obtained with the European Southern Observatory Very Large Telescope, Paranal, Chile, under Large Program 185.A-0791, and made available by the VUDS team at the CESAM data center, Laboratoire d'Astrophysique de Marseille, France. Some of the data presented herein were obtained at Keck Observatory, which is a private, non-profit organization operated as a scientific partnership among the California Institute of Technology, the University of California, and the National Aeronautics and Space Administration. The Observatory was made possible by the generous financial support of the W.M.Keck Foundation. This work was also based on the public SC4K sample of LAEs \citep{sc4k}. The authors acknowledge the hard work of all members of each respective program used to validate spectroscopically the predictions. Finally, we are grateful for the publicly available programming language PYTHON, including the packages: Pandas \citep{pandas}, Numpy \citep{numpy}, Matplotlib \citep{matplotlib}, Seaborn \citep{seaborn}, and scikit-learn \citep{scikit-learn}.

\end{acknowledgements}

\bibliographystyle{aa}

\appendix

\onecolumn\section{Magnitude distribution}

\begin{figure*}[h]
\centering
\includegraphics[width=0.9\textwidth]{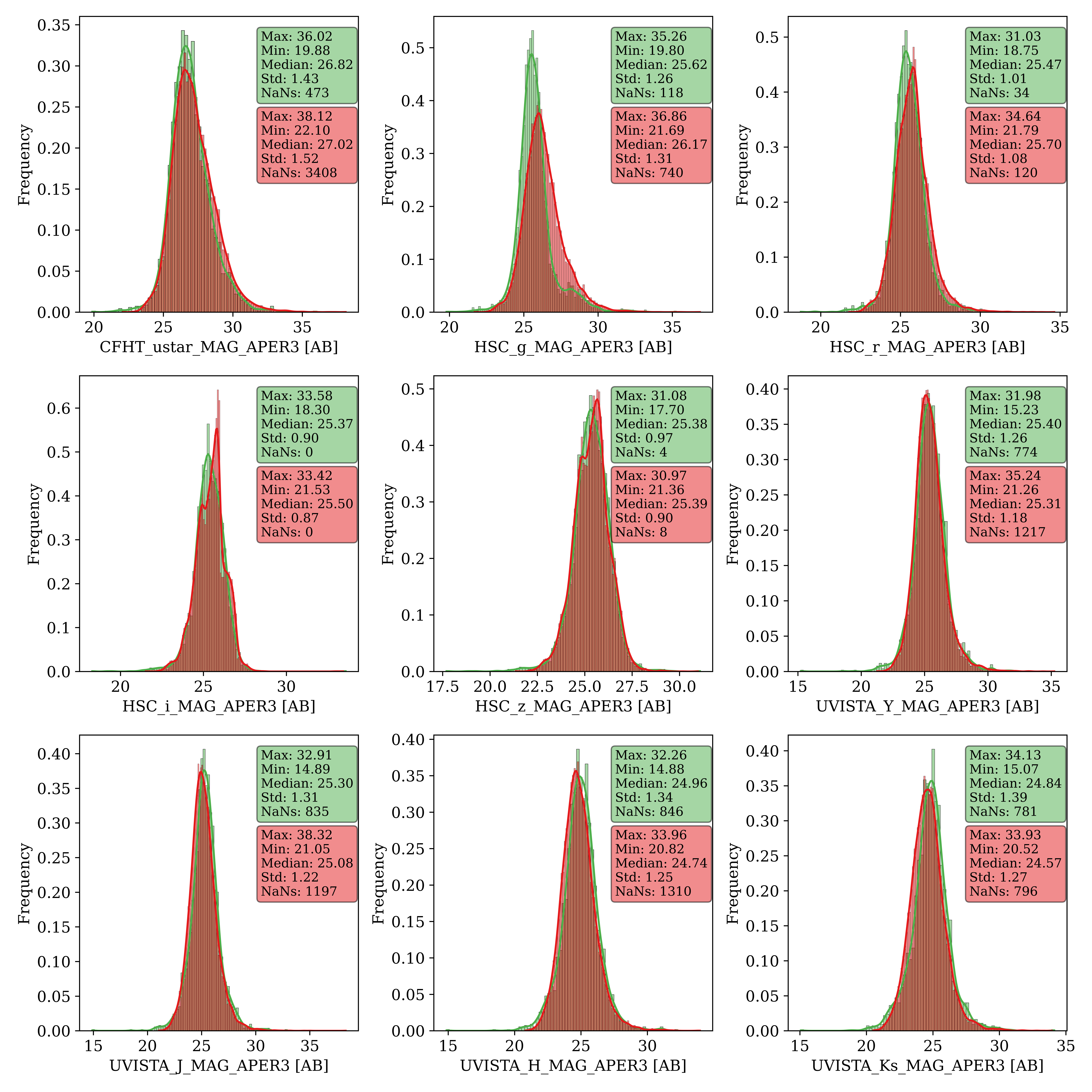}
\caption[nLAE and LAE samples magnitudes distribution and statistics.]{nLAE and LAE samples magnitudes distribution and statistics (maximum, minimum, median, standard deviation, and number of NaNs values). The LAE and nLAE samples are colored green and red, respectively.}
\label{mag_distribution}
\end{figure*}

In this appendix, we present the distribution of magnitudes, along with key statistics such as maximum, minimum, median, standard deviation, and the count of null values. This is presented separately for the LAE sample and the five nLAE subsamples in Figure \ref{mag_distribution}.

\onecolumn\section{Hyperparameter optimization grid}
\label{appendix_tune}

We present the grid used to perform the hyper-parameter optimization (Table \ref{table:optimization_grid}), together with a description of every hyper-parameter taken into consideration for optimization. In Table \ref{final_hyper-parameter_set} the best hyper-parameter set found for each algorithm and sample, ranked by F1-Score from GridSearch, is presented. \par

There are three hyper-parameters with a highlighted importance: \textit{num\_{estimators}} (number of boosting iterations), \textit{learning\_{rate}} (gradient step size), \textit{depth} (maximum depth of each trained tree), and \textit{subsample} (percentage of rows used per iteration to fit each tree).\par

\begin{table}[!h]
\caption{Grids used in the hyper-parameter optimization procedure.}
\centering
\normalsize
\begin{tabular}{|c|c|c|}
\hline
{ \textbf{Algorithm}} & {\textbf{Default hyper-parameters}}  & { \textbf{Optimization grid}}        \\ \hline
\textbf{LightGBM}            & \begin{tabular}[c]{@{}c@{}}n\_estimators = 100\\ learning\_rate = 0.1\\ max\_depth = -1\end{tabular}      & \begin{tabular}[c]{@{}c@{}}n\_estimators = \{50,100,250,500,750,1000\}\\ learning\_rate = \{0.5,0.3,0.1,0.075,0.05,0.02,0.01\}\\ max\_depth = \{-1,4,5,6,7,8\}\end{tabular} \\ \hline
\textbf{XGBoost}         & \begin{tabular}[c]{@{}c@{}}n\_estimators = 100\\ learning\_rate = 0.3\\ max\_depth = 6\end{tabular} & \begin{tabular}[c]{@{}c@{}}n\_estimators = \{50,100,250,500,1000,1500,2000\}\\ learning\_rate = \{0.5,0.3,0.2,0.1,0.075,0.05,0.02,0.01\}\\ max\_depth = \{4,5,6,7,8\}\end{tabular} \\ \hline
\textbf{CatBoost}        & \begin{tabular}[c]{@{}c@{}}n\_estimators = 1000\\ learning\_rate = 0.03\\ max\_depth = 6\end{tabular}                              & \begin{tabular}[c]{@{}c@{}}n\_estimators = \{250,500,600,700,800,900,1000,1250,2000\}\\ learning\_rate = \{0.5,0.3,0.1,0.075,0.05,0.03,0.01\}\\ max\_depth = \{4,5,6,7,8\}\end{tabular} \\ \hline
\end{tabular}

\tablefoot{The values for max\_depth of 0 or -1 result in not limited trees.}
\label{table:optimization_grid}
\end{table}

\begin{table}[ht]
\normalsize
\caption{Final set of hyper-parameters for each algorithm and each sample, selected by the best F1-Score from GridSearch.}
\centering
\begin{tabular}{lcccc}
\toprule
& \multicolumn{4}{c}{Hyper-parameters} \\
\cmidrule(lr){3-5}
Algorithm & Sample & n\_estimators & learning\_rate & max\_depth \\
\midrule
{LightGBM} & Sample 1 & 750 & 0.075 & 5 \\
& Sample 2 & 100 & 0.3 & 7 \\
& Sample 3 & 50 & 0.1 & 8 \\
& Sample 4 & 500 & 0.05 & 8 \\
& Sample 5 & 250 & 0.1 & 7 \\
\addlinespace
{XGBoost} & Sample 1 & 250 & 0.1 & 6 \\
& Sample 2 & 250 & 0.3 & 8 \\
& Sample 3 & 100 & 0.5 & 6  \\
& Sample 4 & 1500 & 0.01 & 7 \\
& Sample 5 & 250 & 0.075 & 4 \\
\addlinespace
{CatBoost} & Sample 1 & 1500 & 0.05 & 4 \\
& Sample 2 & 2000 & 0.1 & 5  \\
& Sample 3 & 1000 & 0.075 & 5 \\
& Sample 4 & 100 & 0.05 & 5 \\
& Sample 5 & 250 & 0.1 & 6 \\
\bottomrule
\end{tabular}\relax
\label{final_hyper-parameter_set}
\end{table}

\onecolumn\section{Confusion matrices separated by redshift interval}
\label{appendix_z_impact}

\begin{figure}[h!]
    \centering
    \begin{subfigure}[b]{1\textwidth}
        \includegraphics[width=\textwidth]{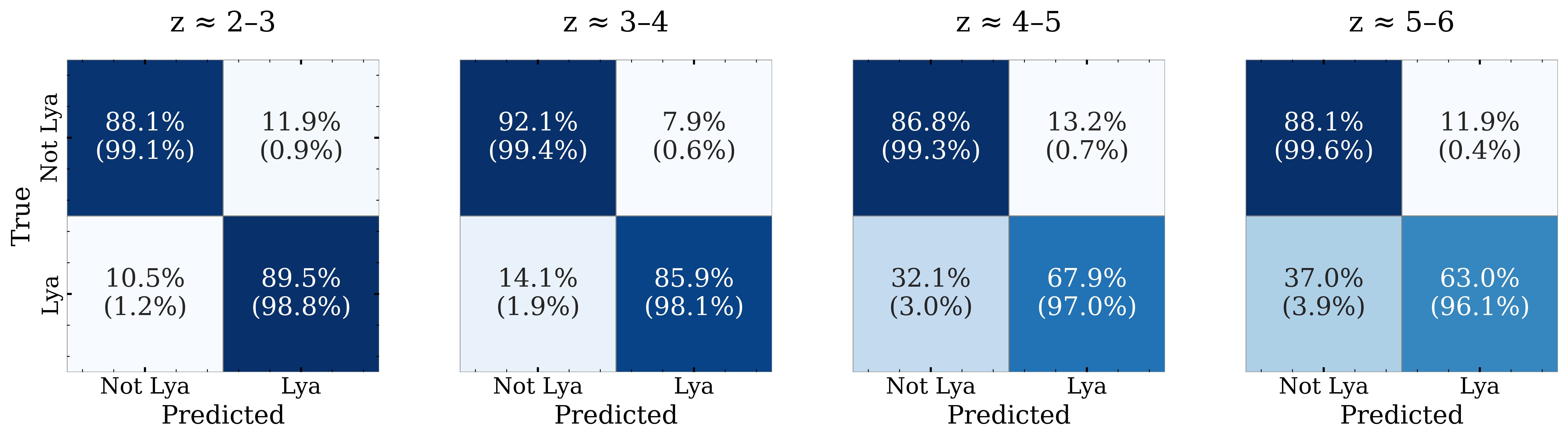}
        \caption{}
        \label{fig:cm_lgbm}
    \end{subfigure}
    
    \vspace{1em}
    
    \begin{subfigure}[b]{1\textwidth}
        \includegraphics[width=\textwidth]{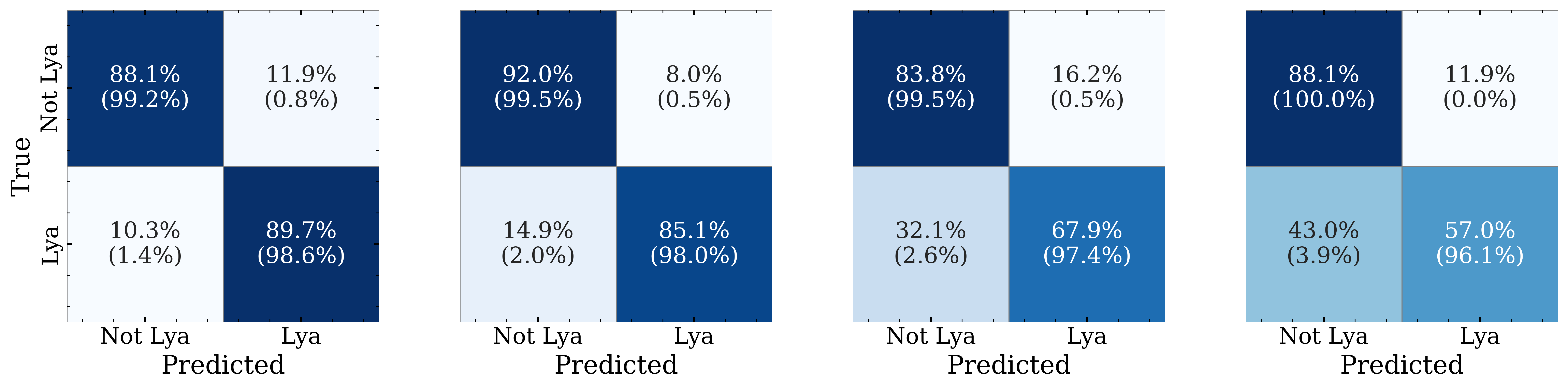}
        \caption{}
        \label{fig:cm_xgb}
    \end{subfigure}
    
    \vspace{1em}
    
    \begin{subfigure}[b]{1\textwidth}
        \includegraphics[width=\textwidth]{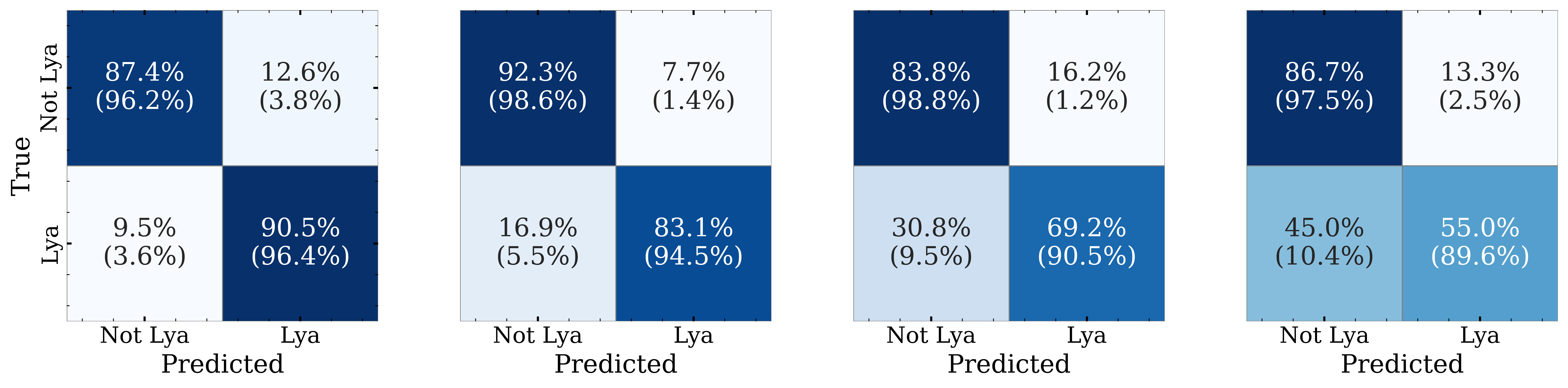}
        \caption{}
        \label{fig:cm_cb}
    \end{subfigure}
    
    \caption{Average confusion matrices for the training (in brackets) and test sets across different redshift intervals using LightGBM (a), XGBoost (b), and CatBoost (c).}
    \label{fig:cm_all_models}
\end{figure}

In this appendix, we present the average confusion matrices for the training and test sets across different redshift intervals: $[2,3], [3,4], [4,5]$, and $[5,6]$. This is presented in Figure \ref{fig:cm_all_models}, separately for each algorithm used. We verify that the number of true positives and true negatives is reduced as the redshift increases, leading to a poorer performance. This effect is more noticeable at $z>4$ and is consistent across the three algorithms.

\onecolumn\section{Spectroscopically confirmed predicted LAEs}
\label{spectroscopic_cosmos}

\begin{longtable}{cccccc}
\caption{Identification of the spectroscopically confirmed LAEs that were predicted by our models. The source compilation used was \citet{cosmos_compilation}. For each source, we present their Cosmos ID, the COSMOS2020 right ascension and declination in degrees, the spectroscopic redshift from the corresponding survey, the average prediction score attributed by our models and the corresponding reference of the observing program.}\label{spec_conf}\\

\toprule
ID & RA & DEC & $z$ & Prediction score & Reference \\
\midrule
\endfirsthead

\multicolumn{6}{c}%
{{ \textbf{Table \thetable.\ }Continued.}} \\
\toprule
ID & RA & DEC & $z$ & Prediction score & Reference \\
\midrule
\endhead

\midrule
\multicolumn{6}{r}{{Continued on next page}} \\
\endfoot

\bottomrule
\endlastfoot
234035 & 149.93480 & 1.63806 & 4.270 & 0.64 & \citet{deimos} \\
\midrule
364107 & 150.28945 & 1.76708 & 3.536 & 0.93 & \citet{deimos} \\
\midrule
882240 & 149.96797 & 2.25817 & 4.825 & 0.83 & \citet{deimos} \\
\midrule
1015255 & 150.34350 & 2.38053 & 4.865 & 0.66 & \citet{deimos} \\
\midrule
1063117 & 149.96235 & 2.42651 & 4.331 & 0.53 & \citet{deimos} \\
\midrule
1065656 & 149.86276 & 2.42895 & 4.252 & 0.92 & \citet{deimos} \\
\midrule
1145690 & 150.32957 & 2.50640 & 4.375 & 0.55 & \citet{deimos} \\
\midrule
1364946 & 149.90115 & 2.71935 & 4.417 & 0.75 & \citet{deimos} \\
\midrule
935357 & 150.13882 & 2.30700 & 3.002 & 0.51 & \citet{vuds_1,vuds_2} \\
\midrule
926044 & 150.10998 & 2.29865 & 2.579 & 0.52 & \citet{vuds_1,vuds_2} \\
\midrule
1027774 & 150.11771 & 2.39110 & 3.685 & 0.68 & \citet{vuds_1,vuds_2} \\
\midrule
986789 & 150.17796 & 2.35369 & 2.663 & 0.76 & \citet{vuds_1,vuds_2} \\
\midrule
1186357 & 150.19970 & 2.54533 & 3.99 & 0.80 & \citet{vuds_1,vuds_2} \\
\midrule
1194293 & 150.17779 & 2.55310 & 2.931 & 0.86 & \citet{vuds_1,vuds_2} \\
\midrule
1189474 & 150.11301 & 2.54828 & 2.475 & 0.75 & \citet{vuds_1,vuds_2} \\
\midrule
960881 & 150.12266 & 2.33115 & 3.100 & 0.83 & \citet{gto_muse} \\
\midrule
855981 & 150.11998 & 2.23466 & 5.245 & 0.52 & \citet{gto_muse} \\
\midrule
851264 & 150.16004 & 2.23055 & 3.362 & 0.50 & \citet{gto_muse} \\
\midrule
840426 & 150.10451 & 2.22100 & 5.306 & 0.77 & \citet{gto_muse} \\
\midrule
835220 & 150.16006 & 2.21626 & 3.009 & 0.69 & \citet{gto_muse} \\
\midrule
833340 & 150.12719 & 2.21475 & 3.393 & 0.96 & \citet{gto_muse} \\
\midrule
830930 & 150.17496 & 2.21229 & 2.973 & 0.53 & \citet{gto_muse} \\
\midrule
830219 & 150.09378 & 2.21195 & 3.393 & 0.80 & \citet{gto_muse} \\
\midrule
828510 & 150.10382 & 2.21038 & 3.244 & 0.94 & \citet{gto_muse} \\
\midrule
820571 & 150.12293 & 2.20322 & 4.97 & 0.86 & \citet{gto_muse} \\
\midrule
841033 & 150.09003 & 2.22157 & 3.247 & 0.81 & \citet{smuvs} \\
\midrule
862699 & 150.15868 & 2.24061 & 3.613 & 0.81 & \citet{smuvs} \\
\midrule
963145 & 150.11921 & 2.33318 & 3.006 & 0.63 & \citet{smuvs} \\
\midrule
297187 & 149.64846 & 1.70356 & 5.755 & 0.56 & \citet{ning_2020} \\
\midrule
897108 & 150.10767 & 2.27187 & 5.858 & 0.77 & \citet{candelsz7} \\
\midrule
709082 & 149.91040 & 2.10059 & 2.602 & 0.98 & \citet{clamato_1,clamato_2} \\
\midrule
746299 & 149.91977 & 2.13525 & 2.414 & 0.73 & \citet{clamato_1,clamato_2} \\
\midrule
746990 & 149.89142 & 2.13569 & 2.885 & 0.99 & \citet{clamato_1,clamato_2} \\
\midrule
772090 & 150.36777 & 2.15830 & 2.614 & 0.87 & \citet{clamato_1,clamato_2} \\
\midrule
779372 & 150.12211 & 2.16507 & 2.986 & 0.61 & \citet{clamato_1,clamato_2} \\
\midrule
780988 & 150.26258 & 2.16598 & 2.601 & 0.77 & \citet{clamato_1,clamato_2} \\
\midrule
818249 & 150.20273 & 2.20032 & 2.628 & 0.99 & \citet{clamato_1,clamato_2} \\
\midrule
831491 & 150.03077 & 2.21253 & 2.710 & 0.98 & \citet{clamato_1,clamato_2} \\
\midrule
1005589 & 150.22114 & 2.37094 & 2.517 & 0.71 & \citet{clamato_1,clamato_2} \\
\midrule
1014763 & 149.98645 & 2.37882 & 2.545 & 0.96 & \citet{clamato_1,clamato_2} \\
\midrule
1072458 & 149.89117 & 2.43519 & 2.682 & 0.70 & \citet{clamato_1,clamato_2} \\
\midrule
1087513 & 149.95876 & 2.45026 & 2.627 & 0.99 & \citet{clamato_1,clamato_2} \\
\midrule
1107108 & 150.03117 & 2.46885 & 2.540 & 0.83 & \citet{clamato_1,clamato_2} \\
\midrule
1115349 & 150.33797 & 2.47537 & 2.403 & 0.67 & \citet{clamato_1,clamato_2} \\
\midrule
1127921 & 150.14896 & 2.48828 & 2.380 & 0.60 & \citet{clamato_1,clamato_2} \\
\midrule
1001949 & 150.12686 & 2.36819 & 2.675 & 0.99 & \citet{hetdex_2} \\
\midrule
882398 & 150.19893 & 2.25778 & 2.611 & 0.99 & \citet{hetdex_2} \\
\midrule
876293 & 150.27823 & 2.25241 & 2.611 & 0.99 & \citet{hetdex_2} \\
\midrule
902799 & 150.24821 & 2.27721 & 2.880 & 0.97 & \citet{hetdex_2} \\
\midrule
861537 & 150.22581 & 2.23847 & 2.947 & 0.95 & \citet{hetdex_2} \\
\midrule
919305 & 150.26890 & 2.29190 & 2.763 & 0.92 & \citet{hetdex_2} \\
\midrule
920040 & 150.11345 & 2.29203 & 2.289 & 0.91 & \citet{hetdex_2} \\
\midrule
852094 & 150.11932 & 2.23086 & 2.886 & 0.88 & \citet{hetdex_2} \\
\midrule
1000124 & 150.26582 & 2.36609 & 2.552 & 0.86 & \citet{hetdex_2} \\
\midrule
869170 & 150.24103 & 2.24515 & 3.444 & 0.84 & \citet{hetdex_2} \\
\midrule
846881 & 150.20089 & 2.22575 & 2.481 & 0.84 & \citet{hetdex_2} \\
\midrule
904304 & 150.11306 & 2.27872 & 2.138 & 0.80 & \citet{hetdex_2} \\
\midrule
875173 & 150.09854 & 2.25180 & 3.173 & 0.78 & \citet{hetdex_2} \\
\midrule
975976 & 150.20227 & 2.34395 & 3.326 & 0.68 & \citet{hetdex_2} \\
\midrule
817254 & 150.14991 & 2.19838 & 2.698 & 0.50 & \citet{hetdex_2} \\
\bottomrule
\end{longtable}

\begin{figure*}[h]
\centering
\includegraphics[width=1.\textwidth]{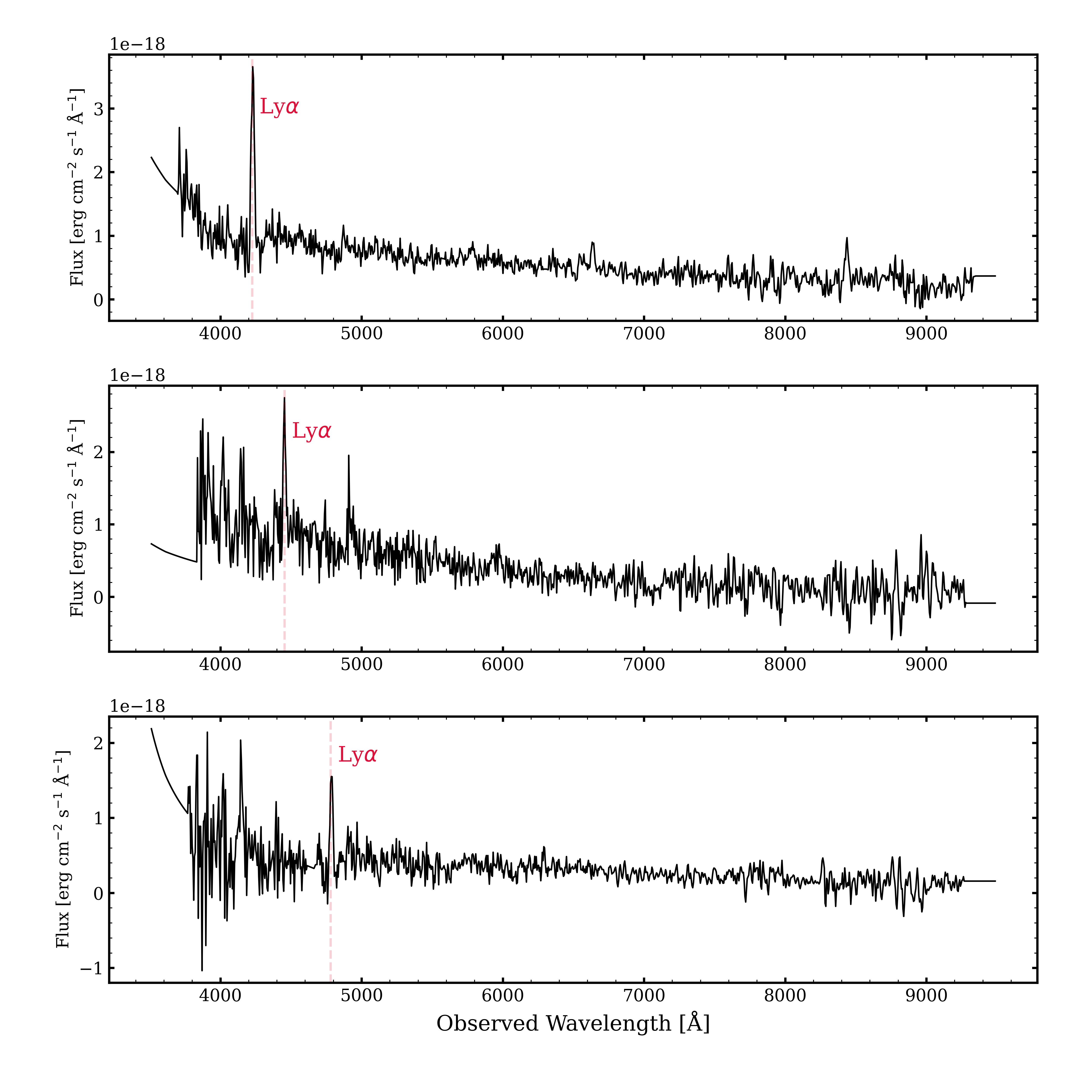}
\caption{Validation spectra for three sources from Table \ref{spec_conf} with IDs 1189474 (top), 986789 (middle), and 1194293 (bottom).}
\label{validation_plot_example}
\end{figure*}

\onecolumn\section{Feature importance}
\label{feature_importances}

\begin{figure}[h]
    \centering
    \begin{subfigure}{0.5\textwidth}
        \includegraphics[width=\linewidth]{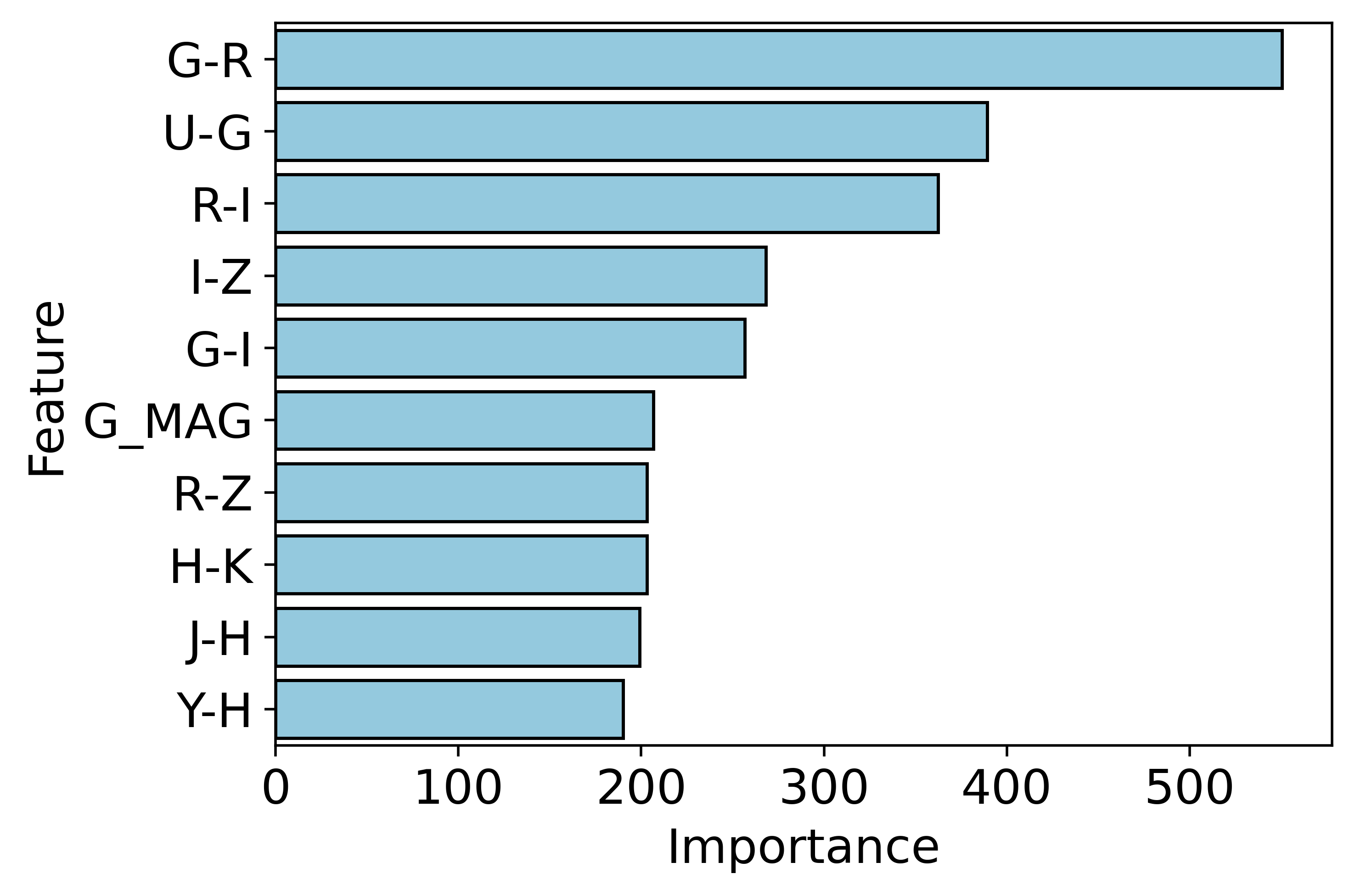}
    \end{subfigure}
    \begin{subfigure}{0.5\textwidth}
        \includegraphics[width=\linewidth]{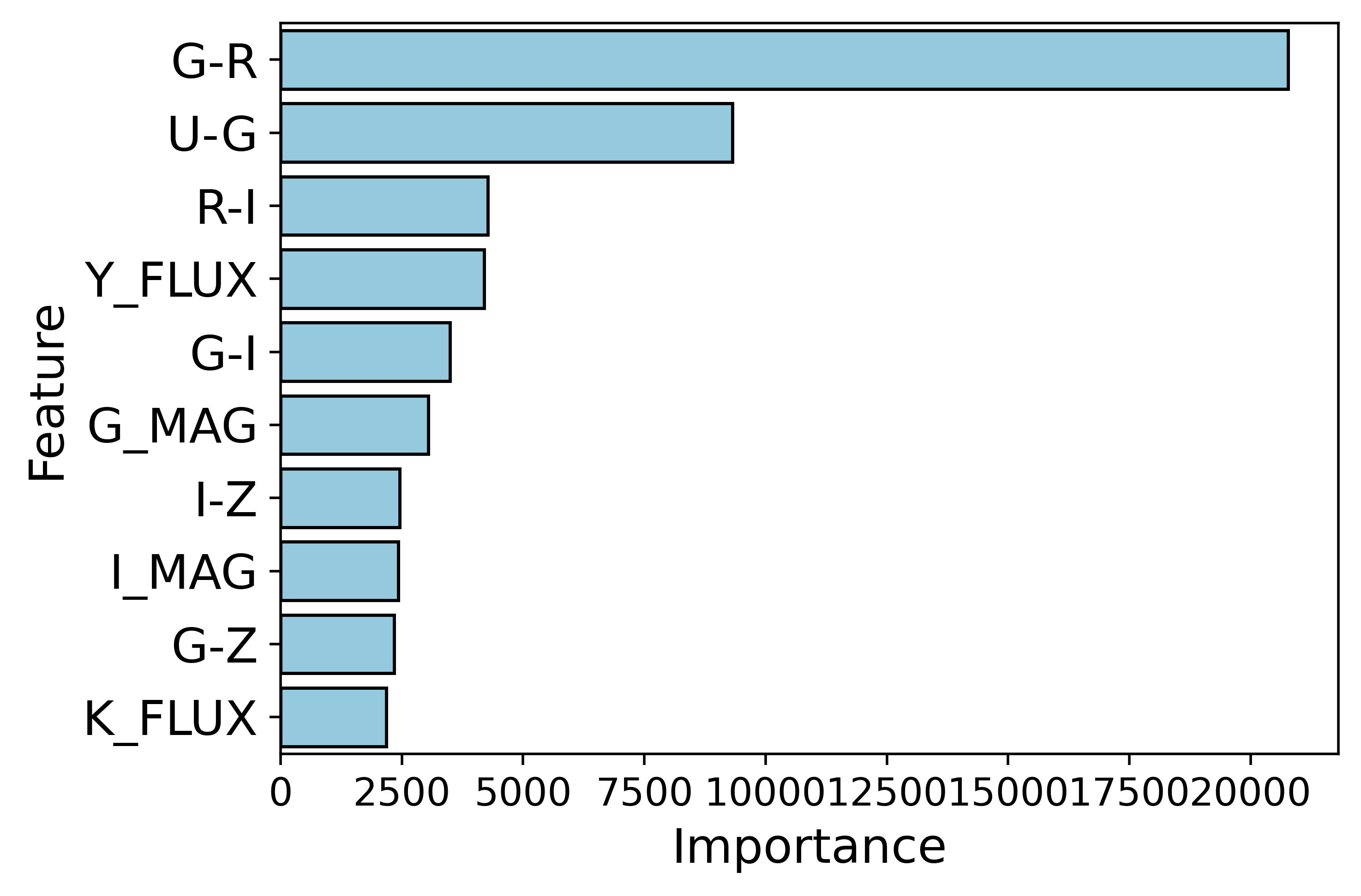}
    \end{subfigure}
    \hfill
    \begin{subfigure}{0.5\textwidth}
        \includegraphics[width=\linewidth]{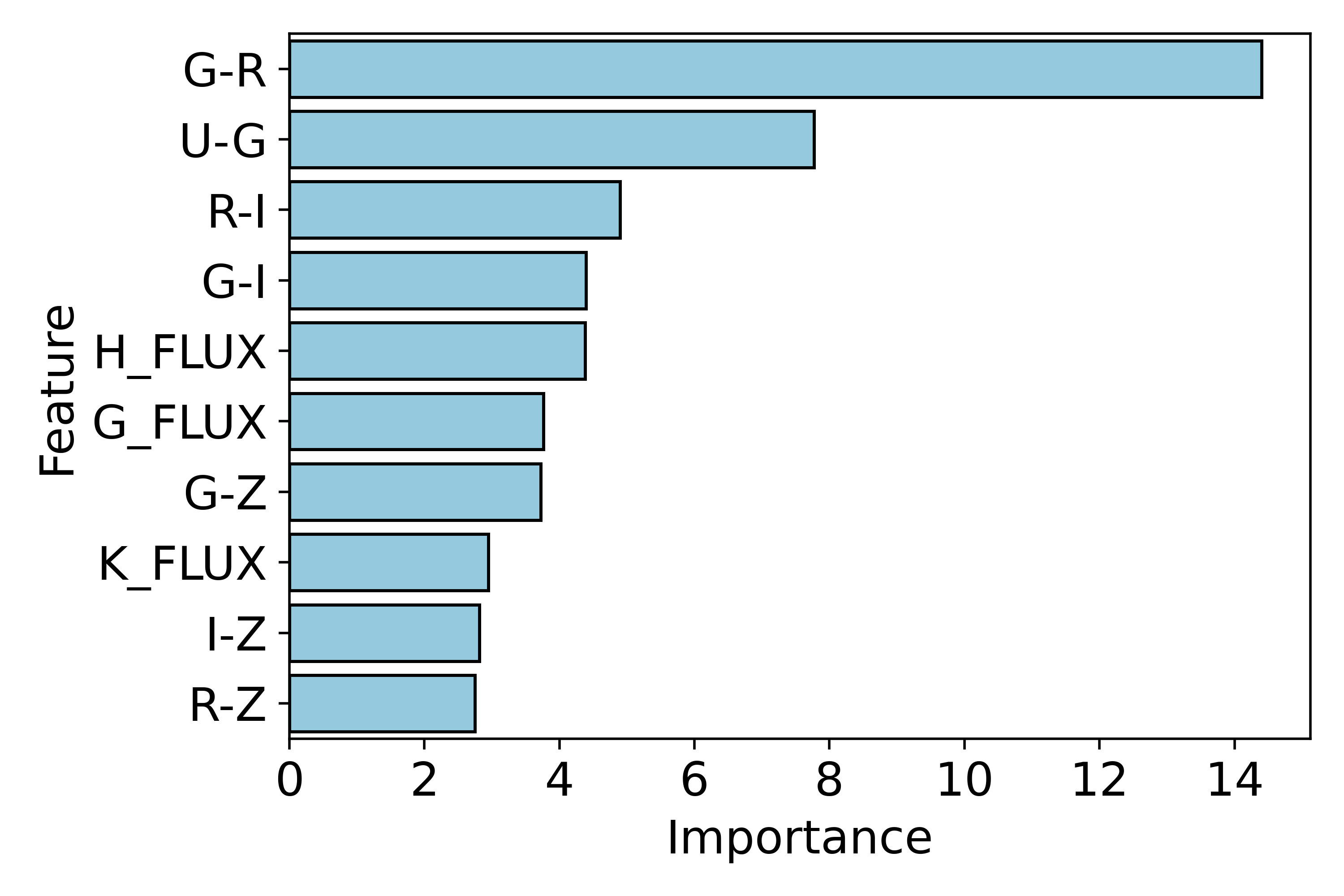}
    \end{subfigure}
    \caption{Average of the feature importances of the 10 most relevant features for LightGBM (top), XGBoost (middle), and CatBoost (bottom).}
    \label{fp_ident}
\end{figure}

In this appendix, we present the ten most important features, averaged across the five samples, for all three algorithms (Figure \ref{fp_ident}). Feature importance is the process by which algorithms assign importance to each feature, allowing the user to obtain insights about the most important features for the models to make the decisions. It is assessed differently across the models. LightGBM and XGBoost use a split-based method, while CatBoost relies on how much a prediction changes when a feature value is altered.

\end{document}